\documentclass[compsoc,conference,a4paper,10pt,times]{IEEEtran}
\IEEEoverridecommandlockouts
\usepackage{cite}
\usepackage{amsmath,amssymb,amsfonts}
\usepackage{algorithmic}
\usepackage{graphicx}
\usepackage{textcomp}
\usepackage{bmpsize}
\usepackage{xcolor}
\usepackage{lipsum}
\usepackage[utf8]{inputenc}
\usepackage{pgffor}
\usepackage{multirow}
\usepackage{listings}
\usepackage{amsmath}
\usepackage{xcolor}
\usepackage{tabularx,booktabs}
\usepackage{textcomp}
\usepackage{caption}
\usepackage{adjustbox}
\usepackage{subcaption}
\usepackage[linesnumbered,ruled,vlined]{algorithm2e}
\usepackage{tikz}
\usepackage{scalefnt}
\usetikzlibrary{shapes}
\usetikzlibrary{arrows}
\usetikzlibrary{positioning}
\usepackage{enumitem}
\usepackage{color}
\usepackage{rotating}
\usepackage[colorlinks=true,urlcolor=black]{hyperref}
\usepackage{makecell}
\usepackage{url}
\usepackage{hyperref}
\def\BibTeX{{\rm B\kern-.05em{\sc i\kern-.025em b}\kern-.08em
    T\kern-.1667em\lower.7ex\hbox{E}\kern-.125emX}}
\definecolor{codegreen}{rgb}{0,0.6,0}
\definecolor{codegray}{rgb}{0.5,0.5,0.5}
\definecolor{backcolour}{rgb}{0.95,0.95,0.92}
\definecolor{highlight}{rgb}{0.79,30,30}
\lstdefinestyle{codeSnippet}{
    backgroundcolor=\color{backcolour},   
    commentstyle=\color{codegreen},
    keywordstyle=\color{blue},
    numberstyle=\tiny\color{codegray},
    stringstyle=\color{codepurple},
    basicstyle=\ttfamily\footnotesize,
    breakatwhitespace=false,         
    breaklines=true,                 
    captionpos=b,                    
    keepspaces=true,                 
    numbers=left,                    
    numbersep=5pt,                  
    showspaces=false,                
    showstringspaces=false,
    showtabs=false,                  
    tabsize=2
}

\lstdefinestyle{ASMstyle}
{
	frame = tb,
  belowskip=.4\baselineskip,
  aboveskip=.4\baselineskip,
  	showstringspaces = false,
              	breaklines = true,
  	breakatwhitespace = true,
  	tabsize = 3,
  	numbers = left,
    stepnumber = 1,
    numberstyle = \tiny\color{gray},
    backgroundcolor=\color{white},
    alsoletter={.\$\%},
    basicstyle={\ttfamily\color{black}},
    stringstyle={\ttfamily\color{string-color}},
    keywordstyle={\ttfamily\color{blue}},
    keywordstyle=[2]{\ttfamily\color{green}},
    keywordstyle=[3]{\ttfamily\color{red}},
    keywordstyle=[4]{\ttfamily\color{violet}},
    keywordstyle=[5]{\ttfamily\color{purple}}, 
    keywords = {cmova, cmovae, cmovb, cmove, cmovne, cmovbe, mov, xor, or, jne, add, jmp,  cmp, shl, and, sar, lfence, clt,  and, lea, victim_function, .B1.1, .L2, .B1.2, .B1.3, .L0},
    morekeywords = {},
    alsoletter= \%().,
    morekeywords = [3]{xorb, \%al},
}

\lstdefinestyle{ASMstyle2}
{
	frame = tb,
  belowskip=.4\baselineskip,
  aboveskip=.4\baselineskip,
  	showstringspaces = false,
              	breaklines = true,
  	breakatwhitespace = true,
  	tabsize = 3,
  	numbers = left,
    stepnumber = 1,
    numberstyle = \tiny\color{gray},
    backgroundcolor=\color{white},
    alsoletter={.\$\%},
    basicstyle={\ttfamily\color{black}},
    stringstyle={\ttfamily\color{string-color}},
    keywordstyle={\ttfamily\color{blue}},
    keywordstyle=[2]{\ttfamily\color{green}},
    keywordstyle=[3]{\ttfamily\color{red}},
    keywordstyle=[4]{\ttfamily\color{black}},
    keywordstyle=[5]{\ttfamily\color{purple}}, 
    keywords = {cmova, cmovae, cmovb, cmove, cmovne, cmovbe, mov, xor, or, jne, add, jmp,  cmp, shl, and, sar, lfence, clt,  and, lea, victim_function, .B1.1, .L2, .B1.2, .B1.3, .L0,array2(\%rdi)},
    morekeywords = {},
    alsoletter= \%().,
    morekeywords = [3]{},
}

\lstdefinestyle{ASMstyleCMOVXCHGSET}
{
	frame = tb,
  belowskip=.4\baselineskip,
  aboveskip=.4\baselineskip,
  	showstringspaces = false,
              	breaklines = true,
  	breakatwhitespace = true,
  	tabsize = 3,
  	numbers = left,
    stepnumber = 1,
    numberstyle = \tiny\color{gray},
    backgroundcolor=\color{white},
    alsoletter={.\$\%},
    basicstyle={\ttfamily\color{black}},
    stringstyle={\ttfamily\color{string-color}},
    keywordstyle={\ttfamily\color{blue}},
    keywordstyle=[2]{\ttfamily\color{green}},
    keywordstyle=[3]{\ttfamily\color{red}},
    keywordstyle=[4]{\ttfamily\color{black}},
    keywordstyle=[5]{\ttfamily\color{purple}}, 
    keywords = {cmova, cmovae, cmovb, cmove, cmovbe, mov, xor, or, jne, add, cmp, shl, and, sar, lfence, clt,  and, lea, victim_function, .B1.1, .L2, .B1.2, .B1.3, .L0,array2(\%rdi)},
    morekeywords = {},
    alsoletter= \%().,
    morekeywords = [3]{cmovne,xchg,seta},
}

\lstdefinestyle{ASMcaner}
{
	frame = tb,
  belowskip=.4\baselineskip,
  aboveskip=.4\baselineskip,
  	showstringspaces = false,
              	breaklines = true,
  	breakatwhitespace = true,
  	tabsize = 3,
  	numbers = left,
    stepnumber = 1,
    numberstyle = \tiny\color{gray},
    backgroundcolor=\color{white},
    alsoletter={.\$\%},
    basicstyle={\ttfamily\color{black}},
    stringstyle={\ttfamily\color{string-color}},
    keywordstyle={\ttfamily\color{blue}},
    keywordstyle=[2]{\ttfamily\color{green}},
    keywordstyle=[3]{\ttfamily\color{red}},
    keywordstyle=[4]{\ttfamily\color{violet}},
    keywordstyle=[5]{\ttfamily\color{purple}}, 
    keywords = {cmova, cmovae, cmovb, cmove, cmovne, cmovbe, mov, xor, or,  jne, add, jmp,  cmp, shl, and, sar, lfence, clt,  and, lea, victim_function, .B1.1, .L2, .B1.2, .B1.3, .L0},
    morekeywords = {},
    alsoletter= \%().,
    morekeywords = [3]{movb, andb, .L1},
}

\lstdefinelanguage
   [x64]{Assembler}     % add a "x64" dialect of Assembler
   [x86masm]{Assembler} % based on the "x86masm" dialect
   {morekeywords={victim_function}} % etc.
                  
\begin{document}

\title{FastSpec: Scalable Generation and Detection of Spectre Gadgets \\Using Neural Embeddings}


%\author{Anonymous Submission}

\author{\IEEEauthorblockN{M. Caner Tol}
\IEEEauthorblockA{Worcester Polytechnic Institute}
\IEEEauthorblockA{mtol@wpi.edu}
\and
\IEEEauthorblockN{Berk Gulmezoglu}
\IEEEauthorblockA{Iowa State University}
\IEEEauthorblockA{bgulmez@iastate.edu}
\and
\IEEEauthorblockN{Koray Yurtseven}
\IEEEauthorblockA{Worcester Polytechnic Institute}
\IEEEauthorblockA{kcyurtseven@wpi.edu}
\and
\IEEEauthorblockN{Berk Sunar}
\IEEEauthorblockA{Worcester Polytechnic Institute}
\IEEEauthorblockA{sunar@wpi.edu}}

\maketitle
% Page numbering
%\thispagestyle{plain}
%\pagestyle{plain}
%

\begin{abstract}
Several techniques have been proposed to detect vulnerable Spectre gadgets in widely deployed commercial software. Unfortunately, detection techniques proposed so far rely on hand-written rules which fall short in covering subtle variations of known Spectre gadgets as well as demand a huge amount of time to analyze each conditional branch in software. Moreover, detection tool evaluations are based only on a handful of these gadgets, as it requires arduous effort to craft new gadgets manually.

In this work, we employ both fuzzing and deep learning techniques to automate the generation and detection of Spectre gadgets. We first create a diverse set of Spectre-V1 gadgets by introducing perturbations to the known gadgets. Using mutational fuzzing, we produce a data set with more than 1~million Spectre-V1 gadgets which is the largest Spectre gadget data set built to date. Next, we conduct the first empirical usability study of Generative Adversarial Networks (GANs) in the context of assembly code generation without any human interaction. We introduce SpectreGAN which leverages masking implementation of GANs for both learning the gadget structures and generating new gadgets. This provides the first scalable solution to extend the variety of Spectre gadgets. 

Finally, we propose FastSpec which builds a classifier with the generated Spectre gadgets based on a novel high dimensional Neural Embeddings technique (BERT). For the case studies, we demonstrate that FastSpec discovers potential gadgets with a high success rate in OpenSSL libraries and Phoronix benchmarks. Further, FastSpec offers much greater flexibility and time-related performance gain compared to the existing tools and therefore can be used for gadget detection in large-scale software.  
\end{abstract}


%%%%%%%%%%%%%%%%
\section{Introduction}
%%%%%%%%%%%%%%%%

The new era of microarchitectural attacks began with newly discovered Spectre~\cite{kocher2019spectre} and Meltdown~\cite{Lipp2018meltdown} attacks, which may be exploited to exfiltrate confidential information through microarchitectural channels during speculative and out-of-order executions.
The Spectre attacks target vulnerable code patterns called gadgets, which leak information during speculatively executed instructions. 
While the initial variants of Spectre~\cite{kocher2019spectre} exploit conditional and indirect branches, Koruyeh et al.~\cite{spectrereturns} propose another Spectre variant by poisoning the entries in Return-Stack-Buffers (RSBs). Moreover, new Spectre-type attacks~\cite{chen2019sgxpectre,spectrereturns} are implemented against the SGX environment and even remotely over the network~\cite{schwarz2019netspectre}. These attacks show the applicability of Spectre attacks in the wild.

Unfortunately, chip vendors try to patch the leakages one-by-one with microcode updates rather than fixing the flaws by changing their hardware designs. Therefore, developers rely on automated malware analysis tools to eliminate mistakenly placed Spectre gadgets in their programs. The proposed detection tools mostly implement taint analysis~\cite{wang2018oo7} and symbolic execution~\cite{wang2019kleespectre,guarnieri2020spectector} to identify potential gadgets in benign applications. However, the methods proposed so far are associated with two shortcomings: (1) the low number of Spectre gadgets prevents the comprehensive evaluation of the tools, (2) time consumption exponentially increases when the binary files become larger.
%\begin{enumerate}[itemsep=0pt,parsep=0pt, leftmargin=3mm]
%\item they are based on hand-written rules limiting generalization,
%\item tools were constructed based on only 15 examples written by Kocher~\cite{kocher2018spectre}, 
%\item it takes days to analyze large binary files~\cite{wang2018oo7} and, 
%%\item unsupported instructions/registers weaken Spectre detection~\cite{guarnieri2020spectector}.
%\item unsupported instructions/registers in~\cite{guarnieri2020spectector}.
%\end{enumerate}
Thus, there is a need for a robust and fast analysis tool that can automatically discover potential Spectre gadgets in large-scale commercial software.

%In recent years, Deep Neural Networks (DNNs) proved that advanced learning algorithms have evolved to achieve superhuman performance in challenging tasks such as classifying images better than humans~\cite{krizhevsky2012imagenet}, outperforming players in games~\cite{dota2,Deepmind}, human-like real-time conversational speech~\cite{conversational} and so on. These achievements prove the extraordinary potential of DNNs, which automate the time-consuming tasks that were purely dependent on human effort. %While Convolutional Neural Networks (CNNs) were developed for visual tasks, 
Natural Language Processing (NLP) techniques are applied to automate challenging natural language and text processing tasks ~\cite{radford2019language}. %which consist of two main categories. %The first one involves gathering and generating information where billions of samples are collected all over the internet and a large-scale NLP model is trained to generate new information~. Second, 
Later, NLP techniques have been used in the security domain, such as network traffic~\cite{radford2018sequence} and vulnerability analysis~\cite{redmond2018cross}. %After an assembly code is produced from a piece of C code by using a disassembler, the malware snippets are detected statically through NLP techniques. 
Such applications leverage word~\cite{mikolov2013word2vec} or paragraph~\cite{le2014distributed} embedding techniques to learn the vector representations of the text. %NLP-based methods are integrated into malware analysis tools to improve automation in computer security. 
The success of these techniques heavily depends on the large data sets, which ease training scalable and robust NLP models. However, for Spectre, for instance, the number of available gadgets is only $15$, making it crucial to create new Spectre gadgets before building an NLP-based detection tool.

Generative Adversarial Networks (GANs)~\cite{goodfellow2014generative} are a type of generative models, which aim to produce new examples by learning the distribution of training instances in an adversarial setting. Since adversarial learning makes GANs more robust and applicable in real-world scenarios, GANs have become quite popular in recent years with applications ranging from generating images~\cite{wang2018high,odena2017conditional} to text-to-image translation~\cite{reed2016generative}, etc. While the early applications of GANs focused on computer vision, implementing the same techniques in NLP tasks poses a challenge due to the lack of continuous space in the text. Various mathematical GAN-based techniques have been proposed to achieve better success in generating human-like sentences to overcome this obstacle ~\cite{fedus2018maskgan,gulrajani2017improved}. However, it is still unclear whether GANs can be implemented in the context of computer security to create application-specific code snippets. Additionally, each computer language has a different structure, semantics, and other features that make it more difficult to generate meaningful snippets for a specific application.

Neural vector embeddings~\cite{mikolov2013word2vec,le2014distributed} used to obtain the vector representations of words have proven extremely useful in NLP applications. Such embedding techniques also enable one to perform vector operations in high dimensional space while preserving the meaningful relations between similar words. Typically, supervised techniques apply word embedding tools as an initial step to obtain the vector embedding of each token and then build a supervised model on top. For instance, BERT~\cite{devlin2019bert} was proposed by the Google AI team, which learns the relations between different words in a sentence by applying a self-attention mechanism~\cite{vaswani2017attention}. BERT has exhibited superior performance compared to previous techniques~\cite{sutskever2014seq2seq,mikolov2010rnnlm} when combined with bi-directional learning. Furthermore, the attention mechanism improves GPU utilization while learning long sequences more efficiently. Recently, BERT-like architectures are shown to be capable of modeling high-level programming languages~\cite{lachaux2020unsupervised, feng2020codebert}. However, it is still unclear whether it will be effective to model a low-level programming language, such as Assembly language, and help build more robust malware detection tools, which is the goal of this paper.

\textbf{Our Contributions} Our contributions are twofold. First, we increase the diversity of Spectre gadgets with the mutational fuzzing technique. We start with 15 examples~\cite{kocher2018spectre} and produce 1 million gadgets by introducing various instructions and operands to the existing gadgets. Then, we propose a GAN-based tool, namely, SpectreGAN, which learns the distribution of 1 million Spectre gadgets to generate new gadgets with high accuracy. The generated gadgets are evaluated from both semantic and microarchitectural aspects to verify their diversity and quality. Furthermore, we introduce novel gadgets that are not detected by state-of-the-art detection tools. %Then, taint analysis (\textit{oo7}~\cite{wang2018oo7}) and symbolic execution (\textit{Spectector}~\cite{guarnieri2020spectector}) based methods are rigorously analyzed with a portion of our data set. We introduce new Spectre gadgets that are not detected by \textit{oo7} and \textit{Spectector} tools and demonstrate the limitations of hand-written rule-based techniques. 

%In the second part, we analyze the applicability of advanced NLP techniques on the Assembly language code snippets that form our gadgets. Initially,  SpectreGAN is built on MaskGAN~\cite{fedus2018maskgan} by adapting it to work for Assembly language code snippets. We also show that SpectreGAN effectively learns the gadget structure in a scalable way and learns to generate new gadgets with a high success rate. 
In the second part, we introduce FastSpec, a high dimensional neural embedding based detection technique derived from BERT, to obtain a highly accurate and fast classifier for Spectre gadgets. We train FastSpec with generated gadgets and achieve a 0.998 Area Under the Curve (AUC) score for OpenSSL libraries in the test phase. Further, we apply FastSpec on Phoronix benchmark tests to show that FastSpec outperforms taint analysis-based and symbolic execution-based detection tools as well as significantly decreases the analysis time. 

In summary,
\begin{itemize}[itemsep=0pt,parsep=0pt,leftmargin=3mm]
    \item We extend 15 base Spectre examples to 1 million gadgets by applying a mutational fuzzing technique,
    %\item We analyze \textit{oo7} and \textit{Spectector} tools with a portion of new data set and demonstrate that hand-written rule-based approaches are not effective against all Spectre gadgets,
    \item We propose SpectreGAN which leverages conditional GANs to create new Spectre gadgets by learning the distribution of existing Spectre gadgets in a scalable way,
    \item We show that both mutational fuzzing and SpectreGAN create diverse and novel gadgets which are not detected by \textit{oo7} and \textit{Spectector} tools,
    \item We introduce FastSpec, which is based on supervised neural word embeddings to identify the potential gadgets in benign applications orders of magnitude faster than rule-based methods.
\end{itemize}

%Generating assembly code for data augmentation with GANs\newline
%Function similarity of assembly code using state of the art embedding techniques \newline
%ML based tool for detecting Spectre gadgets\newline
%Evaluation of previous rule-based tools on a new dataset\newline

%Regardless of the program analysis approach, one needs to have as many different examples as possible to detect the target program behavior in the wild. For rule-based approaches, an expert needs to discover and analyze different possible code snippets to create the rules. Likewise, for the probabilistic approaches, e.g. machine learning algorithms, no expert is needed but still, the models need to be trained with large data sets.

%Why GANs?

\textbf{Outline} The paper is organized as follows: First, the background on transient execution attacks and NLP are given in~\autoref{sec:background}. Then, the related work is given in~\autoref{sec:related_work}. Next, we introduce both fuzzing-based and SpectreGAN generation techniques in~\autoref{sec:detection_tools}. A new Transformer-based detection tool, namely, FastSpec is proposed in~\autoref{sec:NLP_classifier}. Finally, we conclude the paper with discussions and limitations in~\autoref{sec:discussion} and conclusion in~\autoref{sec:conclusion}. 
%%%%%%%%%%%%%
\section{Background} \label{sec:background}
%%%%%%%%%%%%%
%Our work draws from two areas: transient execution attacks with an emphasis on Spectre and from NLP tools which are leveraged to generate and identify Spectre gadgets in Assembly.

\subsection{Transient Execution Attacks}
%Bu cumle introductionta kullanilacak bir cumle{The discovery of transient execution attacks attracted attention from researchers in the last decade.}. 
In order to keep the pipeline occupied at all times, modern CPUs have sophisticated microarchitectural optimizations to predict the control flow and data dependencies, where some instructions can be executed ahead of time in the transient domain. %However, the result of each instruction execution should be committed based on their order in the program. 
However, the control-flow predictions are not 100\% accurate, causing them to execute some instructions wrongly. These instructions cause pipeline flush once they are detected, and their results are never committed. Interestingly, microarchitectural optimizations make it possible to leak secrets. The critical period before the flush is commonly referred to as the transient domain.
%Interestingly, microarchitectural optimizations such as Write Transient Forwarding (WTF) forward these transient results to other microarchitectural components, which makes it possible to leak secrets. The critical time period before the flush happens is commonly referred to as the transient domain.

%In recent years, researchers discovered different types of transient execution attacks and they proposed several attack models and defenses.
%While various new transient execution attacks have been discovered, automated detection tools have been proposed in parallel.
There are two classes of attacks in the transient domain~\cite{Canella2019extended}. The first one is called Meltdown-type attacks~\cite{Lipp2018meltdown, Schwarz2019ZombieLoad, canella2019fallout, stecklina2018lazyfp, vanbulck2018foreshadow, ridl} which exploit delayed permission checks and lazy pipeline flush in the re-order buffer.
%out-of-order execution after a fault or a microcode assist. 
The other class is Spectre-type attacks~\cite{kocher2019spectre, storetoload, kiriansky2018speculative, spectrereturns, Maisuradze_2018} that exploit the speculative execution. As most Meltdown-type attacks are fixed in latest microarchitectures and Spectre-type attacks are still applicable to a wide range of targets~\cite{Canella2019extended}, i.e., Intel, AMD, and ARM CPUs, we focus on Spectre-V1 attacks in this study.
%are orthogonal to our work, it is out of scope in this paper.\textcolor{green}{Neden bunu sectigimizi vurgulamamiz gerekebilir. Meltdownlarin hepsi fix edilmedi ama o zaman neden spectreyi sectik? Cozumu bulunmayan sorun gibi gostermek gerekiyor}

Some researchers proposed new designs requiring a change in the silicon level~\cite{khasawneh2018safespec, koruyeh2019speccfi} while others proposed software solutions to mitigate transient execution attacks~\cite{msvcspectre, retpoline}. Although these mitigations are effective against Spectre-type attacks, most of them are not used because of the drastic performance degradation \cite{speculativeloadhardening}or the lack of support in the current hardware. Hence, Spectre-type attacks are not entirely resolved yet, and finding an efficient countermeasure is still an open problem.

\subsubsection{Spectre}

Since a typical software consists of branches and instruction/data dependencies, modern CPUs have components for predicting conditional branches' outcomes to execute the instructions speculatively. These components are called branch prediction units (BPU), which use a history table and other components to make predictions on branch outcomes. 
\vspace{0.3cm}
\begin{lstlisting}[ backgroundcolor=\color{white},
                    frame=single,
                    xleftmargin=2em,
                    framexleftmargin=1.5em,
                    language=C, 
                    caption= Spectre-V1 C Code, 
                    label={lst:spectre}]
void victim_function(size_t x){
	if(x < size)
		temp &= array2[array1[x] * 512];
}
\end{lstlisting}
`
In Spectre attacks, a user fills the history table with malicious entries such that the BPU makes a misprediction. Then, the CPU executes a set of instructions speculatively. As a result of misprediction, sensitive data can be leaked through microarchitectural components, for instance, by encoding the secret to the cache lines to establish a covert channel. For example, in the Spectre gadget in~\autoref{lst:spectre}, the $2^{nd}$ line checks whether the user input \texttt{x} is in the bound of \texttt{array1}. In a normal execution environment, if the condition is satisfied, the program retrieves $x^{th}$ element of \texttt{array1}, and a multiple of the retrieved value (512) is used as an index to access the data in \texttt{array2}. However, under some conditions, the \texttt{size} variable might not be present in the cache. In such occurrences, instead of waiting for \texttt{size} to be available, the CPU executes the next instructions speculatively. 
%based on the entries in \textsf{BPU} 
To eliminate unnecessary stalls in the pipeline. When \texttt{size} becomes available, the CPU checks whether it made a correct prediction or not. If the prediction was wrong, the CPU rolls back and executes the correct path. Although the results of speculatively executed instructions are not observable in architectural components, the access to the \texttt{array2} leaves a footprint in the cache, making it possible to leak the data through side-channel analysis. 

%There are a number of known Spectre variants: Spectre-V1 (bounds check bypass), Spectre-V2 (branch target injection), Spectre-RSB~\cite{spectrereturns, Maisuradze_2018} (return stack buffer speculation), and Spectre-V4~\cite{storetoload} (speculative store bypass). We focus on Spectre-V1 as our primary goal since there is no effective solution other than inserting fence instructions after conditional branches. On the other hand, Spectre-V2 is mitigated by flushing BTB across context switches. Moreover, unlike Spectre-V2 gadgets, the vulnerable code segment needs to be found to stop the leakage in Spectre-V1.% Thus, a detection tool for finding gadgets is needed.

%Hence, \textsf{Spectre-V1} gadgets pose a greater danger than other variants. Although our techniques are evaluated on \textsf{Spectre-V1}, they can be easily adapted to other variations. %The further details of program analysis techniques for Spectre-V1 are given in \autoref{sec:program_analysis_techniques}.

\subsubsection{Program Analysis Techniques}\label{sec:program_analysis_techniques}

There are two main program analysis techniques that are commonly used to detect Spectre gadgets.

\textbf{Taint Analysis:} Taint analysis tracks outside user-controlled variables that possibly leak any secret data. If the tainted variables are consumed by a new variable in the program flow, the latter is also tainted in the information flow. This technique is commonly used in vulnerability detection~\cite{newsome2005dynamic}, malware analysis~, \cite{bayer2009scalable,yin2007panorama} and web applications~\cite{balzarotti2008saner,nguyen2005automatically} where user input misuses are highly likely. Similarly, in Spectre gadgets, the secret dependent operations after conditional branches are potential secret leakage sources. 
In particular, when the branch decision depends on the user input, the secret is subject to be revealed in the speculative execution state.
In order to detect the Spectre-V1 based leakage in benign programs, the taint analysis technique is used in \textit{oo7}.

\textbf{Symbolic Execution:} Symbolic execution is a technique to analyze the program with symbolic inputs. Each path of the conditional branch is executed symbolically to determine the values, resulting in unexpected bugs. The symbolic execution is applied to detect potential information leakage in benign applications. For instance, \textit{Spectector}~\cite{guarnieri2020spectector} aims to identify the memory and control leaks by supplying symbolic inputs to target functions. While the symbolic execution provides a good understanding of underlying bugs for different input values, it is challenging to apply for large-scale projects due to high resource demand.

\subsection{Natural Language Processing}

\subsubsection{seq2seq Architecture}

Sequence to sequence mapping is a challenging process since the text data set has no numeric values. First, the text data is converted to numeric values with embedding methods~\cite{mikolov2013word2vec,mikolov2013efficient}. Then, a DNN model is trained with vector representations of the text. 

A new approach called seq2seq~\cite{sutskever2014seq2seq} was introduced to model sequence-to-sequence relations. The seq2seq architecture consists of encoder and decoder units. Both units leverage multi-layer Long Short Term Memory (LSTM) structures where the encoder produces a fixed dimension encoder vector. The encoder vector represents the information learned from the input sequence. Then, the decoder unit is fed with the encoder vector to predict the input sequence's mapping sequence. After the end of the sequence token is produced by the decoder, the prediction phase stops. The seq2seq structure is commonly used in chatbot engine~\cite{qiu2017alime} since sequences with different lengths can be mapped to each other.

\subsubsection{Generative Adversarial Networks}\label{sec:GAN}

%The machine learning models that learn the joint probability distribution are called generative models. Generative models learn the probability $P(X,Y) = P(X|Y)\times P(Y)$ and infer the posterior probability distribution $P(Y|X)$ from the  Bayes rule. %\begin{equation}
%     P(Y|X) = \frac{P(X|Y)\times P(Y)}{P(X)}.
%\end{equation}

%Training the generative models using an explicit density function is hard due to the mostly intractable definitions of $P(X)$ and $P(Y|X)$. There are several methods~\cite{salakhutdinov2009deep,bengio2014deep} to estimate the probability density function and train the generative model with an objective function that maximizes the likelihood. On the other hand, 
A specialized method of training generative models was proposed by Goodfellow et al.~, \cite{goodfellow2014generative} called generative adversarial networks (GANs). The generative models are trained with a separate discriminator model under an adversarial setting. 
%There are two ways of training GANs: conditional and unconditional. In the conditional setup, the output of the generator is conditioned on a given input, where one can control the data generated by the generator model. 
%In the unconditional setup, the generator learns a mapping from a noise distribution to the distribution of training data set using the feedback coming from the discriminator model.
In~\cite{goodfellow2014generative}, the training of the generative model is defined as, 
\begin{equation} \label{eq:GAN}
\begin{split}
\min_{G}\max_{D}V(D,G) & = \mathbb{E}_{x\sim p_{data}(x)}[log~D(x)] \\
& + \mathbb{E}_{z\sim p_z(z)}[log(1-D(G(z)))].
\end{split}
\end{equation}
    
In~\autoref{eq:GAN}, the generator $G$ and the discriminator $D$ are trained in such a way that $D$, as a regular binary classifier, tries to maximize its confidence $D(x)$ on real data x, while minimizing $D(G(z))$ on generated samples by the $G$.  At the same time, $G$ tries to maximize the confidence of discriminator $D(G(z))$ on generated samples $G(z)$ and minimize $D(x)$ where $x$ is the real data. 

%\subsubsection{MaskGAN}

MaskGAN~\cite{fedus2018maskgan} is a type of conditional GAN technique to establish a good performance out of traditional GANs. MaskGAN is based on seq2seq architecture with an attention mechanism which improves the performance of the fixed-length encoder vectors. Each time a prediction is made by the decoder unit, a part of the input sequence is used instead of the encoder vector. Hence, each token in the input sequence has a different weight on the decoder output. The main difference of MaskGAN from other GAN-based text generation techniques is the token masking approach, which helps to learn the missing texts in a sequence. For this purpose, some tokens are masked that is conditioned on the surrounding context. This technique increases the chance of generating longer and more meaningful sequences out of GANs.

\subsubsection{Transformer and BERT}
% Transformer
Although recurrent models with attention mechanisms learn the representations of long sequences, attention-only models, namely \textit{Transformer} architectures~\cite{vaswani2017attention}, are shown to be highly effective in terms of computational complexity and performance on long-range dependencies. Similar to \textit{seq2seq} architecture, \textit{the Transformer} architecture consists of an encoder-decoder model. The main difference of \textit{Transformer} is that recurrent models are not used in encoder or decoder units. Instead, the encoder unit is composed of $L$ hidden layers where each layer has a multi-head self-attention mechanism with $A$ attention heads and a fully connected feed-forward network. The input embedding vectors are fed into the multi-head attention, and the output of the encoder stack is formed by a feed-forward network, which takes the output of the attention sub-layer. The decoder unit also has $L$ hidden layers, and it has the same sub-layers as the encoder. In addition to one multi-head attention unit and one feed-forward network, the decoder unit has an extra multi-head attention layer that processes the encoder stack output. To process the information in the sequence order, positional embeddings are used with token embeddings where both embedding vectors have a size of $H$.

% BERT
Keeping the same Transformer architecture, Devlin et al.~\cite{devlin2019bert} introduced a new language representation model called BERT (Bidirectional Encoder Representations from Transformers), which surpasses the state-of-the-art scores on language representation learning. BERT is designed to pre-train the token representation vectors of deep bidirectional Transformers. For a detailed description of the architecture, we refer the readers to~\cite{vaswani2017attention,devlin2019bert}. The heavy part of the training is handled by processing unlabeled data in an unsupervised manner. The unsupervised phase is called~\textit{pre-training}, which consists of masked language model training and next sentence prediction procedures.
The supervised phase is referred to as~\textit{fine-tuning}, where the model representations are further trained with labeled data for a text classification task. Both phases are further explained in detail for Spectre gadget detection model in~\autoref{sec:NLP_classifier}.
%Vanilla seq2seq model has two shortcomings which causes a decline in the performance of sequence to sequence modelling. The first one is the sequential nature of RNNs where hidden states depend on the previous hidden state output. This causes a bottleneck since the dependencies prevent the usage of GPU parallelization. The second drawback is that the encoder vector is not sufficient to carry the input sequence information to decoder unit since RNNs are not good at learning long sequences. To overcome the aforementioned problems, Google AI team published a technique called Transformer~\cite{vaswani2017attention} which gives better performance on language understanding compared to other NLP-techniques. The novelty in Transformer architecture is the attention mechanism which allows the decoder unit to extract the useful information from entire input sequence. The relations among tokens in a sequence are learned by encoding the positions of each token and this information is transferred between encoder and decoder units. The application of Transformer in Spectre gadget detection is further explained in~\autoref{sec:NLP_classifier}.

%\input{sections/problem_definition}
%%%%%%%%%%%%%%%%%%%%%%%%%%%%%%%%%
\section{Related Work}\label{sec:related_work}
%%%%%%%%%%%%%%%%%%%%%%%%%%%%%%

\subsection{Spectre attacks and detectors}
\textbf{Spectre Variations and Covert Channels} In the first Spectre study~\cite{kocher2019spectre}, two variants were introduced. While Spectre-V1 exploits the conditional branch prediction mechanism when a bound check is present, Spectre-V2 manipulates the indirect branch predictions to leak the secret. Next, researchers discovered new variants of Spectre-based attacks. For instance, a variant of Spectre focuses on poisoning Return-Stack-Buffer (RSB) entries with the desired malicious return addresses~\cite{spectrereturns, Maisuradze_2018}. Another variant of Spectre called Speculative Store Bypass~\cite{storetoload} takes advantage of the memory disambiguator's prediction to create leakage. Traditionally, secrets are leaked through cache timing differences. Then, researchers showed that there are also other covert channels to measure the time difference: namely using network latency ~\cite{schwarz2019netspectre}, port contention ~\cite{bhattacharyya2019smotherspectre}, or control flow hijack attack based on return-oriented programming~\cite{mambretti2020bypassing} to leak secret data.

%In the first variants of Spectre, secrets are leaked through cache timing difference, e.g., the results are encoded to the different cache lines and the information is extracted by measuring the latency of cache accesses with \textsf{Flush+Reload}~\cite{yarom2014flush+}. Then, researchers showed that there are also other covert channels to measure the time difference. For instance, Schwarz et al.~\cite{schwarz2019netspectre} showed that network latency can be used to leak the secret. Furthermore, two studies~\cite{bhattacharyya2019smotherspectre} demonstrated the usage of port execution components to leak the secret data. Further, Mambretti et al.~\cite{mambretti2020bypassing} introduced a control flow hijack attack based on return oriented programming to leak secret data.

\textbf{Defenses against Spectre} There are various detection methods for speculative execution attacks. Taint analysis is used in \textit{oo7}~\cite{wang2018oo7} software tool to detect leakages. As an alternative way, the taint analysis is implemented in the hardware context to stop the speculative execution for secret dependent data~\cite{schwarz2019context,yu2019speculative}. The second method relies on symbolic execution analysis. Spectector~\cite{guarnieri2020spectector} symbolically executes the programs where the conditional branches are treated as mispredicted. Furthermore, SpecuSym~\cite{guo2019specusym} and KleeSpectre~\cite{wang2019kleespectre} aim to model cache usage with symbolic execution to detect speculative interference, which is based on Klee symbolic execution engine. Following a different approach, Speculator~\cite{mambretti2019speculator} collects performance counter values to detect mispredicted branches and speculative execution domain. Finally, Specfuzz~\cite{oleksenko2019specfuzz} leverages a fuzzing strategy to test functions with diverse set of inputs. Then, the tool analyzes the control flow paths and determines the most likely vulnerable code snippets against speculative execution attacks.

%First, the execution flow is extracted and then the input from an untrusted source is tainted as a potential vulnerable input with a set of hand-written rules is applied to detect vulnerable Spectre branches in the code. %After marking the untrusted user input, the taint analysis continues to the branches, a load instruction that uses tainted source and a store instruction that stores the outcome of the result to the memory. If those requirements are satisfied and the distance of instructions are inside the speculative execution window, the code is found vulnerable.

%Another proposed method to catch Spectre gadgets is to capture both cache behavior and speculative path that causes leakage using symbolic execution. Both KleeSpectre~\cite{wang2019kleespectre} and SpecuSym~\cite{guo2019specusym} aims to detect Spectre gadgets using this methodology.

%\hyperlink{https://github.com/cdisselkoen/pitchfork}{Pitchfork},SpecFuzz,\hyperlink{https://arxiv.org/pdf/2003.05503.pdf}{Bypassing memory safety mechanisms through speculative control flow hijacks}

\subsection{Binary Analysis with Embedding}
Binary analysis is one of the methods to analyze the security of a program. The analysis can be performed dynamically~\cite{nethercote2007valgrind} by observing the binary code running in the system. Alternatively, the binary can also be analyzed statically~\cite{song2008bitblaze}. 
NLP techniques have been applied to binary analysis in recent years. Mostly, the studies leverage the aforementioned techniques to embed Assembly instructions and registers into a vector space. The most common usage of NLP in the binary analysis is to find the similarities between files. Asm2Vec~\cite{ding2019asm2vec} leverages a modified version of the PV-DM model to solve the obfuscation and optimization issues in a clone search. Zuo et al.~\cite{zuo2018neural} and Redmond et al.~\cite{redmond2018cross} solve the binary similarity problem by NLP techniques when the same file is compiled in different architectures. SAFE~\cite{massarelli2019safe} proposes a combination of skip-gram and RNN self-attention models to learn the embeddings of the functions from binary files to find the similarities. %To evaluate how cyberattacks evolve in time, Attack2Vec~\cite{shen2019attack2vec} introduces an application temporal word embeddings to Intrusion Prevention System.

\subsection{GAN-based Text Generation}

The first applications of GANs were mostly applied to computer vision to create new images such as human faces~\cite{karras2017progressive,karras2019style}, photo blending~\cite{wu2019gp}, video generation~\cite{vondrick2016generating}, and so on. However, text generation is a more challenging task since it is more difficult to evaluate the performance of the outputs. An application~\cite{li2017adversarial} of GANs is in the dialogue generation, where adversarial learning and reinforcement are applied together. SeqGAN~\cite{yu2017seqgan} introduces gradient policy update with Monte Carlo search. LeakGAN~\cite{guo2018long} implements a modified policy gradient method to increase the usage of word-based features in adversarial learning. RelGAN~\cite{nie2019relgan} applies Gumbel-Softmax relaxation for training GANs as an alternative method to gradient policy update. SentiGAN~\cite{wang2018sentigan} proposes multiple generators to focus on several sentiment labels with one multi-class generator. However, to the best of our knowledge, the literature lacks GANs applied to the Assembly code generation. To fill this literature gap, we propose SpectreGAN in~\autoref{sec:SpectreGAN}.

%-Adversarial generation of natural language
%-XLNet: Generalized Autoregressive Pretraining for Language Understanding
%-BERT
%-Release Strategies and the Social Impacts of Language Models\cite{solaiman2019release}

%GAN
%Pre training
%Adversarial GAN

%%%%%%%%%%%%%%%%%%%%%%%%%%%%%%%%%
\section{Gadget Generation}\label{sec:detection_tools} \label{sec:dataset}
%%%%%%%%%%%%%%%%%%%%%%%%%%%%%%%%%

We propose both mutational fuzzing and GAN-based gadget generation techniques to create novel and diverse gadgets. In the following sections, details of both techniques and the diversity analysis of the gadgets are given:

%There are several detection tools proposed by researchers as a defence against the Spectre attack. In this paper, we mainly focus on two of them, namely: \textit{oo7} and Spectector. As these tools are built on a set of hand-written rules to detect Spectre gadgets, they cover all the gadgets written by Kocher~\cite{kocher2018spectre}. However, it is not clear whether they achieve the same success rate at detecting new Spectre gadgets. To analyze both tools, we extend our data set by using mutational fuzzing tools. Our approach consists of the following steps: 

\subsection{Gadget Generation via Fuzzing}\label{sec:gadget_verification} 

We begin with fuzzing techniques to extend the base gadgets to create an extensive data set consists of a million Spectre gadgets in four steps.

\begin{algorithm}
\small %\footnotesize
\DontPrintSemicolon
\SetAlgoLined
%\algo
    \KwIn{An Assembly function $A$, a set of instructions $\mathbb{I}_b$ and sets of registers $\mathbb{R}_b$ for different sizes of $b$}
    \KwOut{A mutated Assembly function $A'$ }
     
    $\mathbb{G} := {\mathbb{R}_b \mapsto \mathbb{I}_b}$\;
    $A' = A$\;
    $\textit{MaxOffset}=length(A)$\;
    \For{1:\textit{Diversity}}{
    \For{Offset=1:\textit{MaxOffset}}{
        \For{1:Offset}{
            $\textit{i}_b \gets random(\mathbb{I})$\;
            $\textit{r}_b \gets random(\mathbb{R}_b | \mathbb{G})$\;
            $\textit{l} \gets random(0:length(A'))$\;
            $Insert(\{\textit{i}_b | \textit{r}_b\},A',l)$\; 
           } 
        \textit{Test boundary check}($A'$)\;
        \textit{Test Spectre leakage}($A'$)\;
        
        }
    }
\caption{Gadget generation using mutational fuzzing}
\label{alg:generation}
\end{algorithm}
\setlength{\textfloatsep}{3pt}% Remove \textfloatsep

\begin{itemize}[leftmargin=8pt]
    \item \textbf{Step 1: Initial Data Set}
    There are 15 Spectre-V1 gadgets written in C by Kocher~\cite{kocher2018spectre} and two modified examples introduced by \textit{Spectector}~\cite{guarnieri2020spectector}. For each example, a different attacker code is written %for the ones which needs different configurations than other gadgets to work correctly and 
    to leak the entire secret data completely in a reasonable time.
    \item \textbf{Step 2: Compiler variants and optimization levels}
    Since our target data set is in assembly code format, each Spectre gadget written in C is compiled into x86 assembly functions using different compilers. We compiled each example with \textit{GCC}, \textit{clang}, and \textit{icc} compilers using \textit{-o0} and \textit{-o2} optimization flags. Therefore, we obtain 6 different assembly functions from each C function with AT\&T syntax. %We eliminated some of the Assembly functions which do not behave as a Spectre gadget, after being compiled with certain compiler and optimization level configurations as it is stated in \cite{guarnieri2020spectector}.
    \item \textbf{Step 3: Mutational fuzzing based generation}
    
    We generated new samples with an approach inspired by mutation-based fuzzing technique~\cite{sutton2007fuzzing} as introduced in Algorithm~\ref{alg:generation}. Our mutation operator is the insertion of random assembly instructions with random operands. For an assembly function $A$ with length $L$, we create a mutated assembly function $A'$. We set a limit on the number of generated samples per assembly function $A$ for each \textit{Offset} value, denoted as \textit{Diversity}. We choose a random instruction $\textit{i}_b$ from the instruction set $\mathbb{I}$, and depending on the instruction format of $\textit{i}_b$; we choose random operands $\textit{r}_b$, which are compatible with the instruction in terms of bit size, $b$.  After proper instruction-operand selection, we choose a random position $l$ in $A'$ and insert $\{\textit{i}_b | \textit{r}_b\}$ into that location. We repeat the insertion process until we reach the~\textit{Offset} value. The randomly inserted instruction and register list are given in Appendix~\ref{sec:instructions_inserted_by_fuzzing}.
    %The mutated Assembly sample $A'$ is then tested if it still has the array boundary-check for a given user input and leaks data with Spectre-V1 attack.
    %
    \item \textbf{Step 4: Verification of the generated gadgets} 
    
    Finally, $A'$ is tested whether it is still a Spectre-V1 gadget or not. There are two verification tests that are applied to the generated functions.
    
    The first verification test is applied to make sure that the function still has the proper array boundary-check for given user inputs. Since random instructions are inserted in random locations in the gadget, a new instruction may alter the flags whose value is checked by the original conditional jump. Once the flags are broken, the secret may be leaked without any speculative execution. To test this case, the PoC Spectre-V1 attacker code~\cite{kocher2019spectre} is modified to supply only out-of-bounds inputs to $A'$, which prevents mistraining the branch predictor. If the secret bytes in the PoC code are still leaked, we conclude that the candidate gadget is broken and exclude it from the pool. %Therefore, we detect the functions that have no boundary-check, performing a boundary-check test by giving out-of-bound values to $A'$. If the gadget still leaks the secret, we exclude the candidate gadget from our data set since it is no longer a Spectre-V1 gadget.
    
    %The second test we apply is to determine the functions that leak secret data through speculative execution. 
    If a generated function $A'$ passes from the first test, we apply the PoC Spectre-V1 attack to the gadget and exclude it if it does not leak the secret data through speculative execution. Additionally, the verification code is modified based on Kocher's examples since each example gadget leaks the secret in a different way. For instance, $4^{th}$ example shifts the user input by 1, which affects the leakage mapping in the cache. Therefore, we modified the PoC code to compile it with the generated gadgets together to leak the whole secret. This process is repeated for each example in Kocher's gadget dataset~\cite{kocher2018spectre}, which yields 16 different verification codes. The secret in the gadgets is only decoded via implementing the Flush+Reload technique. Other microarchitectural side-channels are not in the scope of the verification phase.% The secret decoding is only implemented with cache side-channel since the purpose is to verify whether the secret can be leaked or not. %We believe that our verification method introduces no bias to our generated gadgets since the multiplicand in \textit{array1[x] * 512} is treated as a generic immediate value in~\autoref{sec:NLP_classifier}. 
    
    Other Spectre variants such as SmotherSpectre~\cite{bhattacharyya2019smotherspectre} and NetSpectre~\cite{schwarz2019netspectre} are not in our scope. Hence, the generated gadgets that potentially include SmotherSpectre and NetSpectre variants are not verified with other side-channel attacks. Our verification procedure only guarantees that the extracted gadgets leak secret information through cache side-channel attacks. The verification method can be adjusted to other Spectre variants, which is explained further in~\autoref{sec:discussion}.
    %, we aim for a certain \textit{offset} value
    %ortala uretme suresi
    %\textcolor{green}{Buraya, random instruction eklemenin register clobber yapacagini bildigimizden bahsetmek gerekebilir}
\end{itemize}
At the end of the fuzzing-based generation, we obtained a data set of almost 1.1 million Spectre gadgets\footnote{The attacker codes for each example, the entire data set, SpectreGAN, and FastSpec code are available at \url{https://github.com/vernamlab/FastSpec}}. The overall success rate of the fuzzing technique is 5\% out of compiled gadgets. The generated gadgets are used to train SpectreGAN in the next section.%While 1 million gadgets are used to train SpectreGAN and FastSpec, the remaining samples are used to evaluate the state-of-art detection tools.

\subsection{SpectreGAN: Assembly Code Generation with GANs}\label{sec:SpectreGAN}
%%%%%%%%%%%%%%%%%%%%%%

We introduce SpectreGAN, which learns the fuzzing generated gadgets in an unsupervised way and generates new Spectre-V1 variants from existing assembly language samples. The purpose of SpectreGAN is to develop an intelligent way of creating assembly functions instead of randomly inserting instructions and operands. Hence, the low success rate of gadget generation in the fuzzing technique can be improved further with GANs. 

We build SpectreGAN based on the MaskGAN model, with 1.1 million examples generated in~\autoref{sec:dataset}. Since MaskGAN is originally designed for text generation, we modify the MaskGAN architecture to train SpectreGAN on assembly language. Finally, we evaluate the performance of SpectreGAN and discuss challenges in assembly code generation.

\begin{figure*}
    \centering
    \scalefont{1.50}
    %\scalefont{1.75}
    \begin{adjustbox}{width=0.85\textwidth}
\begin{tikzpicture}[
  hid/.style 2 args={
				    rectangle,% split,
				    minimum width=10mm,
				    minimum height=7mm,
				    rounded corners,
				    %rectangle split horizontal,
				    draw=#2,
				    %rectangle split parts=#1,
				    fill=#2!20,
				    outer sep=0mm,
				    inner sep= 1mm} ,
	dense/.style 2 args={
				    rectangle,
				    minimum width=15mm,
				    rounded corners,
				    draw=#2,
				    fill=#2!20,
				    outer sep=0mm,
				    inner sep= 2mm} 
				    ]
    %\tikzstyle{every node}=[font=\large]
   \usetikzlibrary{fit}
    %%%%%%%%%%%%%%%%%%%%%% draw GENERATOR %%%%%%%%%%%%%%%%%%%%%%
  % draw input nodes
  \foreach \i [count=\step from 1] in {\texttt{pushq},\texttt{\%rbp},\textbf{\texttt{<MASK>}},\textbf{\texttt{<MASK>}},\textbf{\texttt{<MASK>}},\texttt{\%rbp}}
    \node (i\step) at (1.8*\step, -1.1) {\i};
    
  % draw output nodes
  \foreach \t [count=\step from 7] in {\texttt{\%rbp},$\widetilde{x}_{t-1}$,$\widetilde{x}_t$,$\widetilde{x}_{t+1}$,\texttt{\%rbp},\texttt{movb}} {
    \node[align=center] (o\step) at (1+1.8*\step, +1.8) {\t};
  }

  \foreach \t [count=\step from 7] in {\texttt{pushq},\texttt{\%rbp},$\widetilde{x}_{t-1}$,$\widetilde{x}_{t}$,$\widetilde{x}_{t+1}$,\texttt{\%rbp}} {
    \node[align=center] (io\step) at (1+1.8*\step, -1.1) {\t};
  }
  % draw ENCODER 
  \foreach \step [count=\s from 1] in {1,...,6} {
  \node[hid={1}{blue}] (e\step) at (1.8*\step, 0) {$e_G^{\s}$};
    \draw[->,thick] (i\step.north) -> (e\step.south);
    %\draw[->] (e\step.north) -> (h\step.south);
  }
  \foreach \step in {1,...,5}
  {\draw[->,thick] (e\step.east) -> ([xshift=1.8cm]e\step.west);}
 
  % draw DECODER
  \foreach \step [count=\s from 1] in {7,...,12} {
    \node[hid={1}{red}] (d\step) at (1+1.8*\step, 0) {$d_G^{\s}$};

    \draw[->,thick] (d\step.north) -> (o\step.south);
     \draw[->,thick] (io\step.north) -> (d\step.south);
  }  
   \foreach \step in {7,...,11}
  {\draw[->,thick] (d\step.east) -> ([xshift=1.8cm]d\step.west);}
  \draw[->,very thick,dashed] (e6) -> (d7) ;
   
  % MASK connections
  \draw[->] (o8.east)..controls ([xshift=0.7cm]o8.east) and ([xshift=-0.7cm]io9.west)..(io9.west);
  \draw[->] ([xshift=0.1cm]o9.east)..controls ([xshift=1.1cm]o9.east) and ([xshift=-1.1cm]io10.west)..([xshift=-0.0cm]io10.west);
  \draw[->] (o10.east)..controls ([xshift=0.7cm]o10.east) and ([xshift=-0.7cm]io11.west)..(io11.west);
  
  % draw ATTENTION layer for generator
  \draw[fill,gray,fill opacity=0.1,text opacity=1,dashed,rounded corners] ([xshift=-0.5cm,yshift=0.5cm]e1) rectangle ++(22,0.60) node [xshift=3.2cm,yshift=-0.3cm,above of=e2,label=left:\textcolor{black}{Attention Layer}]{} ;%(1.5,0.4)--(25.5,0.4)--(25.5,1.4)--(1.5,1.4)--cycle;
 
  % DENSE layer
   \node[dense={1}{green}, minimum width=10.2cm, inner sep=4] (gdense12) at (18.1, 0.8) {};
  
 \node[draw,rectangle,rounded corners,fit=(gdense12)(e1)(e2)(e3)(e4)(e5)(e6)(d7)(d8)(d9)(d10)(d11)(d12),inner sep=6pt,label={[xshift=-10.3cm,yshift=-0cm]\textbf{Generator}}] (rece) {};
 
 %%%%%%%%%%%%%%%%%%%%%%% draw DISCRIMINATOR %%%%%%%%%%%%%%%%%%%%%%
  % draw input nodes
  \foreach \i [count=\step from 1] in {\texttt{pushq},\texttt{\%rbp},\textbf{\texttt{<MASK>}},\textbf{\texttt{<MASK>}},\textbf{\texttt{<MASK>}},\texttt{\%rbp}}
    \node (id\step) at (1.8*\step, -7.1) {\i};
    
  % draw output nodes
 %\foreach \t [count=\step from 7] in {\texttt{pushq},\texttt{\%rbp},\texttt{movq},\texttt{\%rsp},\texttt{","},\texttt{\%rbp}} {
    %\node[align=center] (od\step) at (1+2*\step, -2.0) {\t};
 % }
  
   \foreach \t [count=\step from 7] in {\texttt{pushq},\texttt{\%rbp},$\widetilde{x}_{t-1}$,$\widetilde{x}_{t}$,$\widetilde{x}_{t+1}$,\texttt{\%rbp}} {
    \node[align=center] (iod\step) at (1+1.8*\step, -7.1) {\t};
  }
  % draw ENCODER 
  \foreach \step [count=\s from 1] in {1,...,6} {
  \node[hid={1}{blue}] (ed\step) at (1.8*\step, -6) {$e_D^\s$};
    \draw[->,thick] (id\step.north) -> (ed\step.south);
    %\draw[->] (e\step.north) -> (h\step.south);
  }
  \foreach \step in {1,...,5}
  {\draw[->,thick] (ed\step.east) -> ([xshift=1.8cm]ed\step.west);}
  
  % draw DECODER
  \foreach \step [count=\s from 1] in {7,...,12} {
    \node[hid={1}{red}] (dd\step) at (1+1.8*\step, -6) {$d_D^\s$};

    %\draw[->,thin] (dd\step.north) -> (od\step.south);
     \draw[->,thick] (iod\step.north) -> (dd\step.south);
  }  
  
   \foreach \step in {7,...,11}
  {\draw[->,thick] (dd\step.east) -> ([xshift=1.8cm]dd\step.west);}

  % draw ATTENTION layer for discriminator
 \draw[fill,gray,fill opacity=0.1,text opacity=1,dashed,rounded corners] ([xshift=-0.5cm,yshift=0.5cm]ed1) rectangle ++(22,3.2)  node [xshift=3.2cm,yshift=2cm,above of=ed2,label=left:\textcolor{black}{Attention Layer}]{} ;%\draw[fill,gray,opacity=0.1,thick,dashed,rounded corners] (1.5,-5.7)--(25.5,-5.7)--(25.5,-2.7)--(1.5,-2.7)--cycle ;
 \node (a_t) [draw,rectangle,minimum height=0.6cm,yshift=0.2cm,above of=ed6]{$a_t$};
 \node (c_t) [draw,rectangle,minimum height=0.6cm,yshift=1.1cm,above of=ed5]{$c_t$};
 \node (h_t) [draw,rectangle,minimum height=0.6cm,yshift=1.7cm,above of=dd9]{$\widetilde{h}_t$};
 \node (h) [above of=dd9,xshift=0.3cm,yshift=0cm]{$h_t$};
 \node (h_s) [above of=ed1,xshift=0.3cm,yshift=0cm]{$\overline{h}_s$};
 \node[align=center,above of=h_t,yshift=0.4cm] (r_t)  {$r_t$};
 
 \foreach \step in{1,...,6}
 {\draw[->,dashed] (ed\step.north)--(a_t.south);
  \draw[->] (ed\step.north)--(c_t.south);
 }
 \draw[->] (dd9.north)--(a_t.south);
 \draw[->] (dd9.north)--(h_t.south);
 \draw[->] (a_t.north)--(c_t.south);
 \draw[->] (c_t.east)--(h_t.south);
 \draw[->] (h_t.north)--(r_t);
 
\draw[->,very thick,dashed] (ed6) -> (dd7) ; 

% dense layer
  \node[dense={1}{green}, minimum width=10.2cm,inner sep=4] (ddense12) at (18.1, -2.55) {};

\node[draw,rectangle,rounded corners,fit=(ddense12)(ed1)(ed2)(ed3)(ed4)(ed5)(ed6)(dd7)(dd8)(dd9)(dd10)(dd11)(dd12),inner sep=6pt,label={[xshift=-9.8cm,yshift=-0cm]\textbf{Discriminator}}] (rece) {};
  
  %%%%%%%%%%%%%%%%%%%%%% draw CRITIC %%%%%%%%%%%%%%%%%%%%%%
   \foreach \step in {7,...,12} {
    \node[hid={1}{red}] (cd\step) at (1+1.8*\step, -8.5) {};

    \draw[->,thick] (iod\step.south) -> (cd\step.north);
     %\draw[->,thick] (cd\step.north) -> (cd\step.south);
  }  
  	
  	\foreach \step in {7,...,11}
  	 {\draw[->,thick] (cd\step.east) -> ([xshift=1.8cm]cd\step.west);}
    \node [below of=cd9,yshift=-0.3cm] (b_t)  {$b_t$};
    \draw[->] (cd9.south) -> (b_t);
   
	\node[dense={1}{green}, minimum width=10.2cm,inner sep=4] (cdense) at (18.1, -9.0) {};
	\node[draw,rectangle,rounded corners,fit=(cdense)(cd7)(cd8)(cd9)(cd10)(cd11)(cd12),inner sep=6pt,label={[xshift=-5.3cm,yshift=-0cm]\textbf{Critic}}] (critic) {};
  
%%%%%%%%%%%%%%%%%%%%% CODE SNIPPET %%%%%%%%%%%%%%%%%%%%%%%
  %\node [align=left](l0) at (-2,0) {$\texttt{pushq \%rbp}$};
  \foreach \line[count=\step from 0] in {$\texttt{pushq \%rbp}$,
  										$\texttt{\textbf{movq \%rsp ,} \%rbp}$,
  										$\texttt{movb \%sil , \%al}$,
  										$\texttt{movq \%rdi , -8 ( \%rbp )}$,
  										$\texttt{movb \%al , -9 ( \%rbp )}$,
  										$\texttt{movq -8 ( \%rbp ) , \%rdi }$,
  										{\hspace*{2.5cm}\vdots}
  										}
  	\node [align=left,anchor=west](l\step) at (-8,-\step*0.8) {\line}; % Change x and y to shift green listbox 
  \node[draw,fill=green,opacity=0.1,inner sep=2mm,rectangle,fit=(l0)(l1)(l2)(l3)(l4)(l5)(l6)] (listbox) {} ;
  \node[draw,thick,rectangle,rounded corners,inner sep=0mm,fit=(l0)(l1),label={[xshift=-1cm,yshift=0.2cm]\textbf{Input Gadget}}] (r) {};
  \draw[->,thick] (r.east)..controls([xshift=2cm]r.east) and ([xshift=-2cm]i1.west) ..(i1.west);
\end{tikzpicture}
\end{adjustbox}
    \vspace{-0.7cm}
    \caption{SpectreGAN architecture. Blue and red boxes represent the encoder and decoder LSTM units, respectively. Green boxes represent the softmax layers. The listed assembly function (AT\&T format) on the left is fed to the models after the tokenization process. The critic model and the decoder part of the discriminator get the same sequence of instructions in the adversarial training.}
    \label{fig:gan}
    %\vspace{-0.4cm}
\end{figure*}
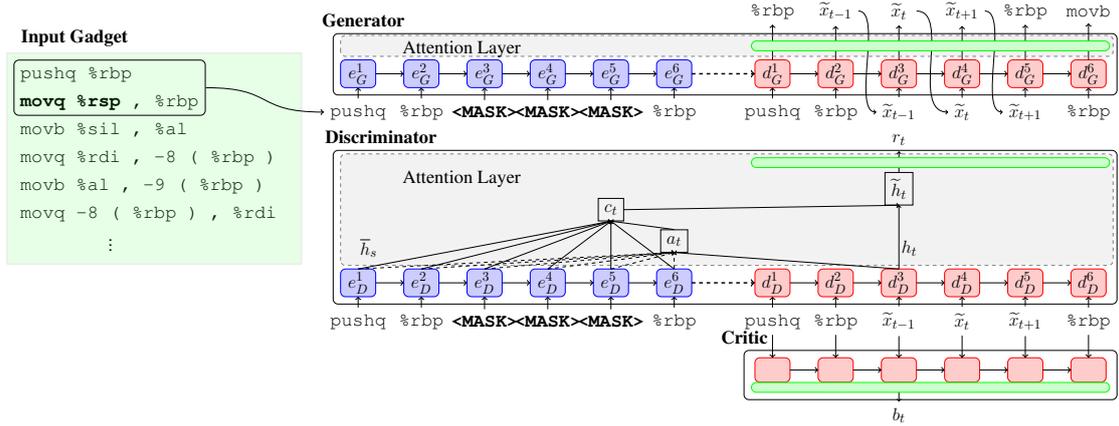
%We define the assembly code generation function as

%We leverage MaskGAN~\cite{fedus2018maskgan} with modifications on the training setup to build SpectreGAN. The generator and discriminator models have seq2seq~\cite{sutskever2014seq2seq} architecture with LSTM-based encoders and decoders with global attention mechanism~\cite{luong2015attention}. Both encoder and decoder units are composed of two-layer stacked LSTM units. For the adversarial training, a critic model is used. Since the assembly functions need to have a correct syntax and instruction order which has a proper start and end, we train our model with padding the shorter assembly functions with \textit{<pad>} token and later mask the loss functions to ignore those paddings. 

\subsubsection{SpectreGAN Architecture}
SpectreGAN has a generator model that learns and generates x86 assembly functions and a discriminator model that gives feedback to the generator model by classifying the generated samples as real or fake as depicted in~\autoref{fig:gan}.

%%%%%%%%%%% Generator Architecture
\textbf{Generator} The generator model consists of encoder-decoder architecture (seq2seq)~\cite{sutskever2014seq2seq}  which is composed of two-layer stacked LSTM units. Firstly, the input assembly functions are converted to a sequence of tokens $T'=\{x'_1,...,x'_N\}$ where each token represents an instruction, register, parenthesis, comma, intermediate value or label. SpectreGAN is conditionally trained with each sequence of tokens where a masking vector $m=(m_1,...,m_N)$ with elements $m_t \in \{0,1\}$ is generated. The masking rate of $m$ is determined as $r_m = \dfrac{1}{N}\sum_{t=1}^N{m_t}$. $m(T')$ is the modified sequence where $x'_t$ is replaced with \texttt{<MASK>} token for the corresponding positions of $m_t=1$. Both $T'$ and $m(T')$ are converted into the lists of vectors $T=\{x_1,...,x_N\}$ and $m(T)$ by a lookup in a randomly initialized embedding matrix of size $V\times H$, where $V$ and $H$ are the vocabulary size and embedding vector dimension, respectively. In order to learn the masked tokens, $T$ and $m(T)$ are fed into the encoder LSTM units of the generator model. Each encoder unit outputs a hidden state $\overline{h}_s$ which is also given as an input to the next encoder unit. The last encoder unit ($e_{G}^{6}$ in~\autoref{fig:gan}) produces the final hidden state which encapsulates the information learned from all assembly tokens.

The decoder state is initialized with the encoder's final hidden state, and the decoder LSTM units are fed with $m(T)$ at each iteration. To calculate the hidden state $\widetilde{h}_t$ of each decoder unit, the attention mechanism output and the current state of the decoder $h_t$ are combined. %encoder hidden states $\overline{h}_s$ and  with the global attention layer~\cite{luong2015attention}. 
The attention mechanism reduces the information bottleneck between encoder and decoder and eases the training~\cite{bahdanau2015attention} on long token sequences in assembly function data set. The attention mechanism is implemented exactly same for both generator and discriminator model which is illustrated in the discriminator part in~\autoref{fig:gan}. The alignment score vector $a_t$ is calculated as:
\begin{equation}\label{eq:alignment}
    a_t(s) =\dfrac{e^{h_t^\top \overline{h}_s}}{\sum_{s'=1}^N{e^{h_t^\top \overline{h}_{s'}}}},
\end{equation}
where $a_t$ describes the weights of $\overline{h}_s$, for a token $x'_t$ at time step $t$, where $h_t^\top \overline{h}_s$ is the score value between the token $x'_t$ and $T'$. This forces decoder to consider the relation between each instruction, register, label and other tokens before generating a new token. The context vector $c_t$ is calculated as the weighted sum of $\overline{h}_s$ as follows:
\begin{equation}\label{eq:context}
    c_t = \sum_{s'=1}^N{a_t(s)\overline{h}_{s'}}.
\end{equation}

For a context vector, $c_t$, the final attention-based hidden state, $\widetilde{h_t}$, is obtained by a fully connected layer with hyperbolic tangent activation function,
\begin{equation}\label{eq:attention}
    \widetilde{h}_t = tanh(W_c[c_t;h_t]),
\end{equation}
where $[c_t;h_t]$ is the concatenation of $c_t$ and $h_t$ with the trainable weights $W_c$. The output list of tokens $\widetilde{T}=(\widetilde{x}_1,...,\widetilde{x}_N)$ is then generated by filling the masked positions for $m(T')$ where $m_t=1$. The probability distribution $p(y_t|y_{1:t-1},x_t)$ is calculated as,
\begin{equation}\label{eq:gen_prob}
    p(y_t|y_{1:t-1},x_t) = \dfrac{e^{W_s\widetilde{h}_t}}{\sum e^{W_s\widetilde{h}_t}},
\end{equation}
where $y_t$ is the output token and attention-based hidden state $\widetilde{h_t}$ is fed into the softmax layer which is represented by the green boxes in~\autoref{fig:gan}. It is important to note that the softmax layer is modified to introduce a randomness at the output of the decoder by a sampling operation. The predicted token is selected based on the probability distribution of vocabulary, \textit{i.e.} if a token has a probability of 0.3, it will be selected with a 30\% chance. This prevents the selection of the token with the highest probability every time. Hence, at each run the predicted token would be different which increases the diversity in the generated gadgets.

%%%%%%%%%%% Discriminator Architecture
\textbf{Discriminator} The discriminator model has a very similar architecture to the generator model. The encoder and decoder units in the discriminator model are again two-layer stacked LSTM units. The embedding vectors $m(T)$ of tokens $m(T')$, where we replace $x'_t$ with \texttt{<MASK>} when $m_t=1$, are fed into the encoder. The hidden vector encodings $\overline{h}_s$ and the encoder's final state are given to the decoder.

The LSTM units in the decoder are initialized with the final hidden state of the encoder and $\overline{h}_s$ is given to the attention layer. The list of tokens $\widetilde{T}$ which represents the generated assembly function by the generator model is fed into the decoder LSTM unit with \textit{teacher forcing}. The previous calculations for $a_t(s)$, $c_t$ and $\widetilde{h}_t$ stated in~\autoref{eq:alignment},~\ref{eq:context}, and~\ref{eq:attention} are valid for the attention layer in the discriminator model as well. The attention-based state value $\widetilde{h}_t$ is fed through the softmax layer which outputs only one value at each time step $t$,
\begin{equation}
    p_D(\widetilde{x}_t=x^{real}_t|\widetilde{T}) = \dfrac{e^{W_s\widetilde{h}_t}}{\sum e^{W_s\widetilde{h}_t}},
\end{equation}
which is the probability of being a real target token $x_t^{real}$.%${y}_t$ where ${y}_t$ is the target actual token. 

SpectreGAN has one more model apart from the generator and the discriminator models, which is called the critic model, and it has only one two-layer stacked LSTM unit. The critic model is initialized with zero states and gets the same input $\widetilde{T}$ with the decoder. The output of the LSTM unit at each time step t is given to the softmax layer, and we obtain
\begin{equation}
    p_C(\widetilde{x}_t=x^{real}_t|\widetilde{T}) = \dfrac{e^{W_b {h}_t}}{\sum e^{W_b {h}_t}},
\end{equation}
which is an estimated version of $p_D$. The purpose of introducing a critic model for probability estimation will be explained in~\autoref{sec:gan_training}.
\subsubsection{Training}\label{sec:gan_training}
The training procedure consists of two main phases namely, pre-training and adversarial training. 

\paragraph{Pre-training phase} The generator model is first trained with maximum likelihood estimation. The real token sequence $T'$ and masked version $m(T')$ are fed into the generator model's encoder. Only the real token sequence $T'$ is fed into the decoder using \textit{teacher forcing} in the pre-training. The training maximizes the log-probability of generated tokens, $\widetilde{x}_t$  given the real tokens, $x'_t$, where $m_t=1$. Therefore, the pre-training objective is
%Given a mask rate $m$, randomly chosen $N \times m$ tokens from T are masked, where $N$ is the number of tokens in $T$. The encoder unit gets one token and decoder unit outputs a probability distribution $P$ over the vocabulary of tokens $V$ at a time step. After the token selection on $P$, the cross-entropy loss is calculated only for the masked token positions. The objective of the generator is minimizing the masked cross-entropy loss function.
%
\begin{equation}
    \dfrac{1}{N} \sum_{t=1}^N{\log{p(m(\widetilde{x}_t)|m(x'_t))}},
\end{equation}
where $p(m(\widetilde{x}_t)|m(x'_t))$ is calculated only for the masked positions. The masked pre-training objective ensures that the model is trained for a \textit{Cloze} task~\cite{taylor1953cloze}.
\paragraph{Adversarial training phase} The second phase is adversarial training, where the generator and the discriminator are trained with the GAN framework. Since the generator model has a sampling operation from the probability distribution stated in~\autoref{eq:gen_prob}, the overall GAN framework is not differentiable. We utilize the policy gradients to train the generator model, as described in the previous works~\cite{yu2017seqgan,fedus2018maskgan}.
%In this phase, we select a contiguous block of tokens of size $N\m$ to be masked, keeping the starting position random, and replace the selected tokens with a special token \texttt{<MASK>}. The masked list of tokens is fed into the encoder of generator as it is shown in Figure~\ref{fig:gan}. After feeding $t_N$, the last token of the assembly function, the state $h_N$ of the encoder is transferred into the decoder of generator and the unmasked tokens are given as inputs. Similarly to pre-training phase, decoder $d_G$ selects the next token from the probability distribution on the $V$. The output of the decoder is fed as an input if the input token is \texttt{<MASK>} in the next time step.

%The new assembly functions constructed by the generator are fed into the encoder of discriminator in the same way with generator's encoder where the masked tokens are replaced by \texttt{<MASK>} token. The state $h_N$ of encoder is again transferred into decoder. The decoder of discriminator D is fed with the fake samples generated by the generator G. At each time step, the output of LSTM unit is given to the fully connected neural network layer with sigmoid activation function which outputs $P(\tilde{x}_t=x^{real}_t|{\tilde{x}}_{0:N})$, which is the probability of being a real token of $\tilde{x}_t$ generated by the generator G. 

The reward $r_t$ for a generated token $\widetilde{x}_t$ is calculated as the logarithm of $p_D(\widetilde{x}_t=x^{real}_t|\widetilde{T})$. The aim of the generator model is to maximize the total discounted rewards $R_t=m(\sum_{s=t}^{N} \gamma^{s} r_s)$ for the fake samples, where $\gamma$ is the discount factor. Therefore, for each token, the generator is updated with the gradient in \autoref{eq:reinforce_gradient} using the REINFORCE algorithm, where $b_t=\log{p_C(\widetilde{x}_t=x^{real}_t|\widetilde{T})}$ is the baseline rewards by the critic model. Subtracting $b_t$ from $R_t$ helps reducing the variance of the gradient~\cite{fedus2018maskgan}. 
%\begin{equation}\label{eq:reinforce_gradient}
%    r_t = log(\dfrac{1}{1+e^{-D(x)}}) %+\epsilon)
%\end{equation}
%\begin{equation}\label{eq:reinforce_gradient}
%    V^*(s_t) = Q^\pi(s_t,a_t) = \mathbb{E}[m(\sum_{s=1}^N\gamma^{s-t} r_s)]
%\end{equation}
%\begin{equation}\label{eq:reinforce_advantage}
%    A^\pi(s_t,a_t) = Q^\pi(s_t,a_t) - V^\pi(s_t)
%\end{equation}
%% discriminator
%% attention luong
\begin{equation}\label{eq:reinforce_gradient}
    \nabla_\theta\mathbb{E}_G[R_t] = (R_t - b_t)\nabla_\theta \log G_\theta(\tilde{x}_t)
\end{equation}

To train the discriminator model, both real sequence $T$ and fake sequence $\widetilde{T}$ are fed into the discriminator. Then, the model parameters are updated such that $\log{p_D(\widetilde{x}_t=x^{real}_t|\widetilde{T})}$ is minimized and $\log{p_D(x_t=x^{real}_t|T)}$ is maximized using maximum log-likelihood estimation.

%In text generation GANs, the evaluation of the generated samples mostly depend on an oracle or an evaluation metric like BLEU score. We use a compiler as an oracle for evaluation which is a direct and intuitive way, since it is not clear if the existing evaluation metrics are suitable for assembly code snippets. Therefore, the test process in Algorithm~\ref{alg:generation} is still applicable for the GAN evaluation as well. 

% TODO: lstm unitleri stacked yap
%

%
\subsubsection{Tokenization and Training Parameters} \label{sec:gan_details}
%\paragraph{\textbf{Pre-processing}} 
Firstly, we pre-process the fuzzing generated data set to convert the assembly functions into sequences of tokens, $T'=(x'_1,...,x'_N)$. We keep commas, parenthesis, immediate values, labels, instruction and register names as separate tokens. To decrease the complexity, we reduce the tokens' vocabulary size and simplify the labels in each function so that the total number of different labels is minimum. %Since we do not insert any immediate value in the mutation operators, no simplification is needed for them. 
The tokenization process converts the instruction "\texttt{movq~(\%rax),~\%rdx}" into the list \texttt{["movq", "(", "\%rax", ")", ",", "\%rdx"]} where each element of the list is a token $x'_t$. Hence, each token list $T'=\{x'_1,...,x'_N\}$ represents an assembly function in the data set. 

The masking vector has two different roles in the training. While a random masking vector $m=(m_1,...,m_N)$ is initialized for the pre-training, we generate $m$ as a contiguous block with a random starting position in the adversarial training. In both training phases, the first token's mask is always selected as $m_1=0$, meaning that the first token given to the model is always real. The masking rate, $r_m$ determines the ratio of masked tokens in an assembly function whose effect on code generation is analyzed further in~\autoref{sec:gan_eval}.

SpectreGAN is configured with the embedding vector size of $d=64$, generator learning rate of $\eta_G=5\times10^{-4}$, discriminator learning rate of $\eta_D=5\times10^{-3}$, critic learning rate of $\eta_C=5\times10^{-7}$ and discount rate of $\gamma=0.89$ based on the  MaskGAN implementation~\cite{fedus2018maskgan}. We select the sequences with a maximum length of 250 tokens, building the vocabulary with a size of $V=419$. We separate 10\% of the data set for model validation. SpectreGAN is trained with a batch size of $100$ on NVIDIA GeForce GTX 1080 Ti until the validation perplexity converges in~\autoref{fig:test}. The pre-training lasts about 50 hours, while the adversarial training phase takes around 30 hours.

\subsubsection{Evaluation} \label{sec:gan_eval}

\begin{figure}[t!]
    \centering
    \includegraphics[width=\columnwidth, height=5cm]{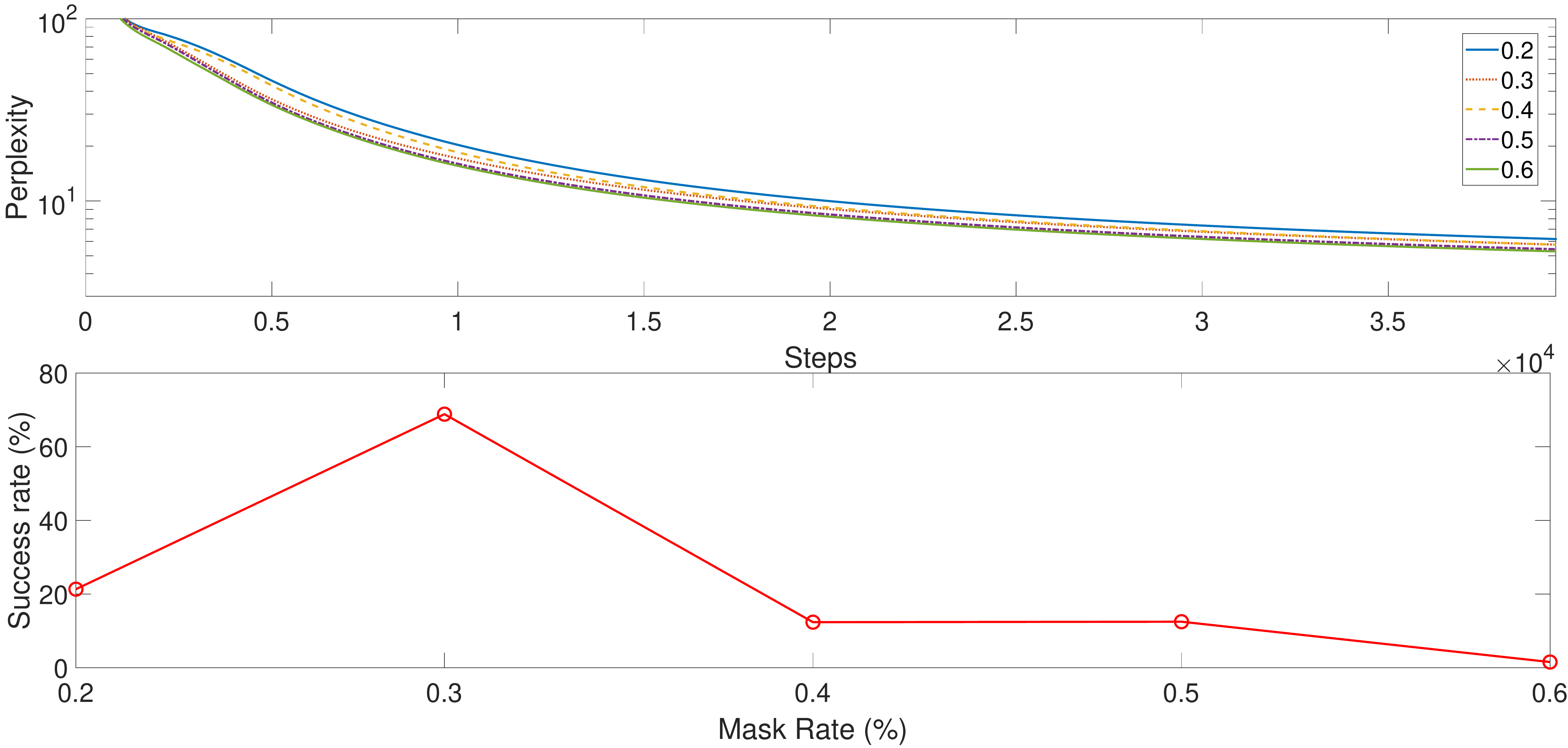}
    \caption{(Above) The validation perplexity decreases at each training step and converges for all $r_m$. (Below) Spectre gadget success rates are evaluated when different masking rates are used to train SpectreGAN. %Compilation success rate (blue) represents the percentage of samples which are compiled without any error out of 100,000 generated functions. 
    Spectre gadget success rate shows the percentage of gadgets out of compiled functions.
    }
    \label{fig:test}
    %\vspace{-0.18cm}
\end{figure}

SpectreGAN is based on learning masked tokens with the surrounding tokens. The masking rate is not a fixed value, which is determined based on the context. Since SpectreGAN is the first study to train on Assembly functions, the masking rate choice is of utmost importance to generate high-quality gadgets. Typically, NLP-based generation techniques are evaluated with their associated perplexity score, which indicates how well the model predicts a token. Hence, we evaluate the performance of SpectreGAN with various masking sizes and their perplexity scores. In~\autoref{fig:test}, the perplexity converges with the increasing number of training steps, which means the tokens are predicted with a higher accuracy towards the end of the training. SpectreGAN achieves lower perplexity with higher masking rates, which indicates that higher masking rates are more preferable for SpectreGAN.

Even though the higher masking rates yield lower perplexity and assembly functions of high quality in terms of token probabilities, our purpose is to create functions which behave as Spectre gadgets. Therefore, as a second test, we generated 100,000 gadgets for 5 different masking rates. Next, we compiled our gadgets with the \textit{GCC} compiler and then tested them with all the attacker code to verify their secret leakage. When SpectreGAN is trained with a masking rate of 0.3, the success rate of gadgets increases by up to 72\%. Interestingly, the success rate drops for other masking rates, demonstrating the importance of masking rate choice. In total, 70,000 gadgets are generated with a masking rate of 0.3 to evaluate the performance of SpectreGAN in terms of gadget diversity in~\autoref{sec:diversity}.

%We use gcc compiler as an oracle for the evaluation, which is a direct and intuitive way since it is not clear if the existing evaluation metrics on text generation, such as BLEU score, are suitable for assembly code snippets. The generated assembly files are fed into the \textit{gcc} compiler and the compilation success and Spectre-V1 leakage rates are used as the main evaluation metrics for SpectreGAN. Therefore, the test process in Algorithm~\ref{alg:generation} is still applicable to the GAN evaluation as well. 

%In order to obtain the best performance, we trained SpectreGAN with different masking rates, $r_m$. Our results show that the perplexity in the validation data set decreases for all of $r_m$ values in the pre-training phase. It also indicates that the large masking rate values achieve lower perplexity. Since the validation perplexity is not the direct evaluation metric of our model, we generated 100,000 samples from each of the masking options and analyzed the compilation rates. We further tested the successfully compiled gadgets using the attacker codes and obtained the ratio of successful Spectre-V1 gadgets. Figure~\ref{fig:test} shows that using the masking rate of $r_m=0.3$ is the optimum since it achieves the maximum gadget generation success rate when the samples with compilation errors are eliminated. We also claim that validation perplexity is not strongly correlated with gadget generation success rate. Therefore, it is not a good measure for generating assembly functions using GANs. 

To illustrate an example of the generated samples, we fed the gadget in~\autoref{lst:spec_gadget} to SpectreGAN and generated a new gadget in~\autoref{lst:gan_gadget}. We demonstrate that SpectreGAN is capable of generating realistic assembly code snippets by inserting, removing, or replacing the instructions, registers, and labels. In the~\autoref{lst:gan_gadget}, the lines that start with the instructions written with red color are generated by SpectreGAN, and they correspond to the masked portion of Spectre-V1 gadget given in~\autoref{lst:spec_gadget}.    %generated the gadget in Listing~\ref{lst:gan_gadget} conditionally on the gadget in Listing~\ref{lst:spec_gadget}.

\begin{minipage}[b]{0.35\columnwidth}
\vspace{3mm}
\centering
\lstset{label=SliceExaple,columns=flexible}
\begin{lstlisting}[style=ASMstyle2,
                    caption=Input Spectre-V1 gadget,backgroundcolor=\color{white},frame=single, 
                    xleftmargin=-0.4em,
                    framexleftmargin=1em,
                    framexrightmargin=-2.5em,
                    numbers=left, label={lst:spec_gadget}, linewidth =3.96cm,
                    basicstyle=\footnotesize]
victim_function:
.cfi_startproc 
movl    size(%rip),%eax
cmpq    %rdi,%rax
jbe     .L0
leaq    array1(%rip),%rax
movzbl  (%rdi,%rax),%eax 
ror     $1,%rsi
shlq    $9,%rax
leaq    array2(%rip),%rcx
movss   %xmm8,%xmm4
movb    (%rax,%rcx),%al 
andb    %al,temp(%rip) 
movd    %xmm1,%r14d 
test    %r15,%rcx 
sbbl    %r13d,%r9d 
.L0:
retq
cmovll  %r8d,%r10d 
.cfi_endproc 
\end{lstlisting}
\end{minipage}
\hspace{-5mm}
\begin{minipage}[b]{0.49\columnwidth}
\centering
\lstset{label=SliceExaple2,columns=flexible}
\begin{lstlisting}[style=ASMcaner,
                    escapechar=!,
                    caption=Generated gadget by SpectreGAN,backgroundcolor=\color{white},frame=single,
                    xleftmargin=2em,
                    framexleftmargin=1em,
                    framexrightmargin=0em,
                    label={lst:gan_gadget}, 
                    linewidth = 3.92cm, basicstyle=\footnotesize]
victim_function:
.cfi_startproc
movl    size(%rip),%eax
cmpq    %rdi,%rax
jbe     .L0
leaq    array1(%rip),%rax
movzbl  (%rdi,%rax),%eax
ror     $1,%rsi
shlq    $9,%rax
movb    array2(%rdi),%al
andb    %al,temp(%rip)
.L1:
andb    %r13b,%al
movb    array2(%rax),%al
andb    %al,temp(%rip)
sbbl    %r13d,%r9d
.L0:
retq
cmovll  %r8d,%r10d
.cfi_endproc

\end{lstlisting}
\end{minipage}

\iffalse
\begin{minipage}[b]{0.5\columnwidth}
\centering
\lstset{label=SliceExaple}
%\lstinputlisting{example_source.c}
\begin{lstlisting}[caption=Input Spectre-V1 gadget,backgroundcolor=\color{white},frame=single, numbers=left, label={lst:spec_gadget}, linewidth = 4cm]
victim_function: 
.cfi_startproc 
movl array1_size(%rip),%eax
cmpq %rdi,%rax
jbe .L0
leaq array1(%rip),%rax
movzbl (%rdi,%rax),%eax 
ror $1,%rsi
shlq $9,%rax
leaq array2(%rip),%rcx
movss %xmm8,%xmm4
movb (%rax,%rcx),%al 
andb %al,temp(%rip) 
movd %xmm1,%r14d 
test %r15,%rcx 
sbbl %r13d,%r9d 
.L0:
retq
cmovll %r8d,%r10d 
.cfi_endproc 

\end{lstlisting}
\label{lis:spec_gadget}
\caption{Example WHILE code snippet with nested scopes.}
\end{minipage}
\hspace{0.5cm}
\begin{minipage}[b]{0.45\columnwidth}
\centering
\begin{lstlisting}[caption=Generated Spectre-V1 gadget by SpectreGAN,backgroundcolor=\color{white},frame=single, numbers=left, label={lst:gan_gadget}, linewidth = 4cm]
victim_function:
.cfi_startproc
movl    array1_size(%rip),%eax
cmpq    %rdi,%rax
jbe     .L0
leaq    array1(%rip),%rax
movzbl  (%rdi,%rax),%eax
shlq    $9,%rax
movb    array2(%rdi),%al
andb    %al,temp(%rip)
.L1:
andb    %r13b,%al
movb    array2(%rax),%al
andb    %al,temp(%rip)
retq
sbbl    %r13d,%r9d
.L0:
retq
cmovll  %r8d,%r10d
.cfi_endproc

\end{lstlisting}
\end{minipage}
\fi

\subsection{Diversity and Quality Analysis of Generated Gadgets}\label{sec:diversity}

In total, 1.2 million gadgets are generated by the mutational fuzzing technique and SpectreGAN. Since the gadgets are derived from existing examples, it is crucial to analyze their diversity and quality. The diversity is measured by syntactic analysis, e.g., counting the number of unique n-grams in gadgets. For the quality metric, we monitor performance counters while the gadgets are executed. 5000 gadgets are randomly selected from each gadget generation technique to perform syntactic and microarchitectural analysis. Furthermore, novel gadgets that are not detected by \textit{oo7}~\cite{wang2018oo7} and \textit{Spectector}~\cite{guarnieri2020spectector} tools are given to show that our gadget generation techniques produce meaningful Spectre-V1 gadgets.

\subsubsection{Syntactic Analysis}

In NLP applications, the diversity of the generated texts is evaluated by counting the number of unique n-grams. The most common metrics for the text diversity are perplexity and BLEU scores that are calculated based on the probabilistic occurrences of n-grams in a sequence. The higher number of n-grams indicates that an NLP model learns the data set distribution efficiently and produces new sequences with high diversity. However, both scores are obtained during the training phase; thus, making it impossible to evaluate the fuzzing generated gadgets since there is no training phase. Instead, we conduct diversity analysis by counting the unique n-grams introduced by fuzzing and SpectreGAN methods after all the gadgets are generated.

The number of unique n-grams in generated gadgets is compared with 17 base examples in \autoref{tab:n-grams}. The unique n-grams are calculated as follows: First, unique n-grams produced by fuzzing are identified and stored in a list. Then, additional unique n-grams introduced by SpectreGAN are noted. Therefore, the unique n-grams generated by SpectreGAN in~\autoref{tab:n-grams} represent the number of n-grams introduced by SpectreGAN, excluding fuzzing generated n-grams. 

\begin{table}[h]
\centering
\caption{Table shows the number of unique n-grams for base gadgets and generated gadgets by fuzzing and SpectreGAN methods. In the last column the total number of unique n-grams are given as well as the increase factor that improves with the increasing n-grams.}
\small
\begin{tabular}{>{\centering\arraybackslash}m{0.03\columnwidth}|>{\centering\arraybackslash}m{0.09\columnwidth}|>{\centering\arraybackslash}m{0.12\columnwidth}|>{\centering\arraybackslash}m{0.2\columnwidth}|>{\centering\arraybackslash}m{0.28\columnwidth}}
\hline \toprule
 \textbf{n} & \textbf{Base}         & \textbf{Fuzzing}        & \textbf{SpectreGAN} & \textbf{Total}\\  \midrule
2         &    2069      &     15,448   & 7,462    & 22,910 ($\times$11)\\ 
3         &    3349      &     181,606  & 91,851   & 273,457 ($\times$82)\\ 
4         &    4161      &     639,608  & 460,317  & 1,099,925 ($\times$264)\\ 
5         &    4747      &     998,279  & 921,519  & 1,919,798 ($\times$404)\\ 
\bottomrule
\end{tabular}
\label{tab:n-grams}
%\vspace{-7mm}
\end{table}

In total, the number of unique bigrams (2-grams) is increased to 22,910 from 2,069, which is more than 10 times raise. While new instructions and registers added by fuzzing improve the gadgets' diversity, SpectreGAN contributes to the gadget diversity by producing unique perturbations. Since the instruction diversity increases drastically compared to base gadgets, the unique 5-grams reach up to almost 2 million, 400 times higher than the base gadgets. The results show that both fuzzing and SpectreGAN span the diversity in the generated gadgets. High diversity in the gadget data set also results in microarchitectural behavior diversity as well as new Spectre-V1 gadgets that were not previously considered during the design process of previous detection mechanisms.

\subsubsection{Microarchitectural Analysis}

Another purpose of gadget generation is to introduce new instructions and operands to create high-quality gadgets. To assess the quality of the gadgets, we analyze gadgets' microarchitectural characteristics. The first challenge is to examine the effects of instructions in the transient domain since they are not visible in the architectural state. After carefully analyzing the performance counters for Haswell architecture, we determined that two counters, namely, $uops\_issued:any$ and $uops\_retired:any$ give an insight into gadgets' microarchitectural behavior. $uops\_issued:any$ counter is incremented every time a $\mu$op is issued, which counts both speculative and non-speculative $\mu$ops. On the other hand, $uops\_retired:any$ counter only counts the executed and committed $\mu$ops, which automatically excludes speculatively executed $\mu$ops. 
\vspace{0.3cm}
\begin{figure}[t]
    \centering
    \includegraphics[width=\columnwidth]{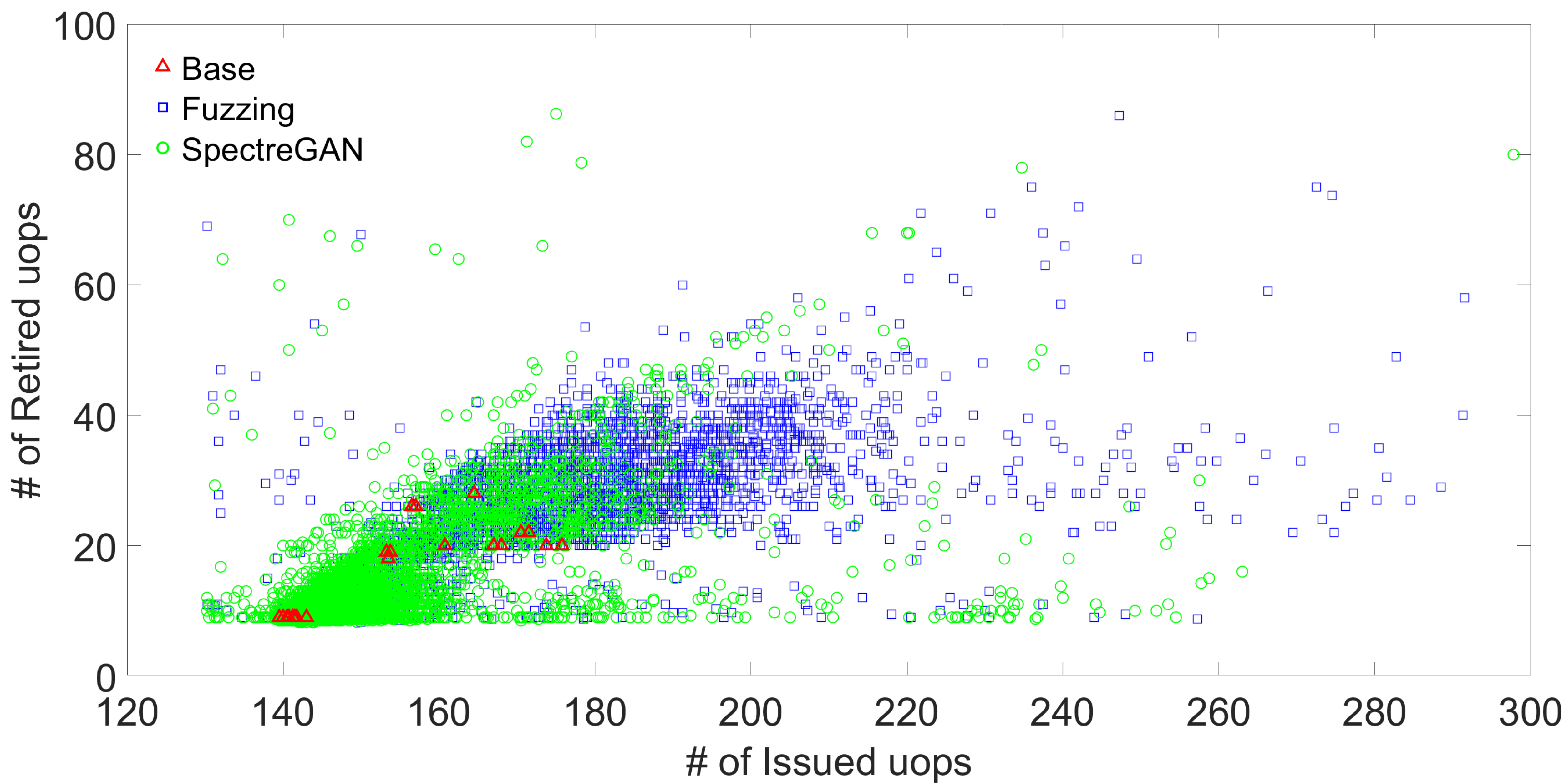}
    \caption{The distribution of base (red-triangle), fuzzing generated (blue-square) and SpectreGAN generated (green-circle) gadgets is given for issued and retired $\mu$ops counters. Both SpectreGAN and fuzzing techniques generate diverse set of gadgets in Haswell architecture.}
    \label{fig:counter_diversity}
    %\vspace{-2mm}
\end{figure}

The performance counter distribution of generated gadgets and base gadgets are given in~\autoref{fig:counter_diversity}. The gadget quality is measured by the number of instructions in the transient domain after a gadget passes the verification step. The exploitable gadgets in the commercial software have many instructions that are speculatively executed until the secret is leaked. If our detection tool in ~\autoref{sec:NLP_classifier} is only trained with simple gadgets from Kocher's examples, the success rate would be low in large-scale software binaries. Moreover, the gadgets that are detected in the case studies are very similar to the generated gadgets which have more instructions in the transient domain. A similar observation is also shared in~\cite{wang2019oo7}, where the authors claim that Spectre gadgets have up to 150 instructions between the conditional branch and speculative memory access in the detected gadgets. Since our aim is to create realistic gadgets by inserting various instructions, we assume that gadget quality increases in parallel when a gadget is close to the x-axis and far from the y-axis. 

It is more likely to obtain high-quality gadgets with fuzzing method as new instructions and operands are randomly added. On the other hand, SpectreGAN learns the essential structure of the fuzzing generated gadgets, which yields almost the same number of samples close to the x-axis in~\autoref{fig:counter_diversity}. Moreover, the advantage of SpectreGAN is to automate the creation of gadgets with a higher accuracy (72\%) compared to the fuzzing technique (5\%).

\subsubsection{Detection Analysis}

Even though the microarchitectural and syntactic analyses show that fuzzing and SpectreGAN can produce diverse and high-quality sets of gadgets, we aim to enable a comprehensive evaluation of detection tools and determine the most interesting gadgets in our data set. For this reason, the generated gadgets are fed into \textit{Spectector}~\cite{guarnieri2020spectector} and \textit{oo7}~\cite{wang2018oo7} tools to determine the novelty of the gadgets. 

\textbf{oo7} tool leverages taint analysis to detect Spectre-V1 gadgets. It is based on the Binary Analysis Platform (BAP)~\cite{BAP} which forwards taint propagation along all possible paths after a conditional branch is encountered. \textit{oo7}~\footnote{https://gitlab.com/igoto/spectre-detector} is built on a set of hand-written rules which cover the existing examples by Kocher~\cite{kocher2018spectre}. Although our data set size is 1.2 million, we have selected 100,000 samples from each gadget example uniformly random due to the immense time consumption of \textit{oo7} (150 hours for ~100K gadgets), which achieves a 94\% detection rate. 

Interestingly, specific gadget types from both fuzzing and SpectreGAN are not caught by \textit{oo7}. When a gadget contains \textit{cmov} or \textit{xchg} or \textit{set} instruction and its variants, it is not identified as a Spectre gadget. Hence, we introduce these gadgets as novel Spectre-V1 gadgets listed in~\autoref{lst:cmov} and~\autoref{lst:prev}. Their corresponding assembly snippets are also given in Appendix~\ref{sec:gadgets_bypass_oo7_spectector}. %When xchg gadget is compiled, {xchg \%rdi, \%r13} instruction stores the content of \%rdi register to \%r13 register. In the next iterations of the secret leakage, the stored value is being accessed. This leakage is not detected since a previous value is used to map the secret the cache lines.  

\vspace{0.3cm}
\begin{lstlisting}[ backgroundcolor=\color{white},
                    frame=single,
                    xleftmargin=2em,
                    framexleftmargin=1.5em,
                    language=C, 
                    caption= {\texttt{CMOV} gadget: An example Spectre gadget in C format. When it is compiled with \textit{gcc-7.5} -o2 optimization level, \texttt{CMOVcc} gadget bypasses \textit{oo7} tool. The generated assembly version is given in Appendix~\ref{sec:gadgets_bypass_oo7_spectector}.}, 
                    label={lst:cmov}]
void victim_function(size_t x){
	if(global_condition)
		x = 0;
	if(x < size)
		temp &= array2[array1[x] * 512];
}
\end{lstlisting}
\vspace{0.1cm}
\begin{lstlisting}[ backgroundcolor=\color{white},
                    frame=single,
                    xleftmargin=2em,
                    framexleftmargin=1.5em,
                    language=C, 
                    caption= {\texttt{XCHG} gadget: When a past value, that is controlled by the attacker, is used to leak the secret in the Spectre gadget, \textit{oo7} cannot detect the XCHG gadget. This example show that control-flow graph extraction is not efficiently implemented in \textit{oo7} tool.}, 
                    label={lst:prev}]
size_t prev = 0xff;
void victim_function(size_t x) {
	if (prev < size)
		temp &= array2[array1[prev] * 512];
	prev = x;
}
\end{lstlisting}

We identified two potential issues of static taint analysis method in \textit{oo7} tool. First, if a portion of a tainted variable is modified by an instruction such as \textit{cmov} or \textit{set}, the tainted variable is not tracked by the tool. However, an attacker still controls the remaining portion of the variable, which makes it possible to leak the secret from memory. In some cases, the implementation of static taint analysis is not sufficiently accurate to track partially modified tainted variables due to under-tainting. Secondly, the tainted variables are not tracked between the iterations of a loop. If an old attacker-controlled variable is used to access the secret, \textit{oo7} tool is not able to taint the old variable between the iterations of a \textit{for} loop. Hence, any old attacker-controlled variable can be used to bypass the tool. This shows that control flow graphs of multiple iterations may not be extracted correctly by \textit{oo7}. Both weaknesses show that hand-written rules do not generalize well for Spectre gadget detection when new Spectre-V1 gadgets are discovered.

\textbf{Spectector}~\cite{guarnieri2020spectector} makes use of a symbolic execution technique to detect the potential Spectre-V1 gadgets. For each assembly file, \textit{Spectector} is adjusted to track 25 symbolic paths of at most 5000 instructions each, with a global timeout of 30 minutes. The remaining parameters are kept as default. 

First, we eliminate the gadgets that include unsupported instructions as these gadgets are never detected by \textit{Spectector}. When we analyze the remaining gadgets, 1\% of the gadgets are not detected successfully. Then, undetected gadgets are examined to determine novel gadgets. %We observed that when the gadgets include either sfence/mfence/lfence or 8-bit registers (\%al, \%bl, \%cl, \%dl) in~\autoref{lst:xorb}, they are likely to bypass \textit{Spectector}. We introduce the corresponding novel gadgets and their code snippets in~\autoref{sec:gadgets_bypass_oo7_spectector}. 

We determined two issues in the \textit{Spectector} tool. The first issue is related to the barrier instructions. Even though \textit{lfence}, \textit{sfence} and \textit{mfence} instructions have different purposes, the tool treats them as equal instructions. For instance, if an \textit{sfence} instruction is present after the conditional branch, the tool classifies the gadget as safe. However, \textit{sfence} instruction has no effect on the load operation so, the gadget still leaks the secret. Hence, Spectector's modeling of fences does not distinguish the differences between various x86 fence instructions. The second issue is about 8-bit registers in which a partial information of the elements in \textit{array[x]} is stored. When 8-bit registers are used to modify the elements in~\autoref{lst:xorb}, \textit{Spectector} is no longer able to detect the gadgets. This second issue is also mentioned in~\cite{guarnieri2020spectector}, i.e., sub-registers are currently not supported by the tool. Overall, these issues are due to the problems in the translation from x86 assembly into Spectector's intermediate language. 

We show that our large-scale diverse gadget data set establishes a ground truth to evaluate the detection tools accurately. As shown in the case studies on \textit{Spectector} and \textit{oo7}, the success rate on detecting the gadgets in our 1.1 million sample data set could serve as a generic evaluation metric while identifying the flaws in the detection tools.

\begin{lstlisting}[style=ASMstyle,
                    frame=single,
                    caption={\textcolor{red}{\textit{xorb \%al, \%al}} is added to $1^{st}$ example in Kocher's examples~\cite{kocher2018spectre}. \textit{Spectector} is no longer able to detect the leakage due to the zeroing \%al register.},
                    label={lst:xorb},
                    xleftmargin=2em,
                    framexleftmargin=1.5em,
                    basicstyle=\footnotesize,
                    float=h]	
victim_function: 	
    movl    size(%rip), %eax	
    cmpq    %rax, %rdi	
    jae     .B1.2
    movzbl  array1(%rdi), %eax	
    shlq    $9, %rax	
    xorb    %al, %al	
    movb    array2(%rax), %dl	
    andb    %dl, temp(%rip)	
.B1.2:	
    ret
\end{lstlisting}

%\input{sections/verifying_gadgets}
%%%%%%%%%%%%%%%%%%%%%%%%%%%%%%%%%
\section{FastSpec: Fast Gadget Detection Using BERT}\label{sec:NLP_classifier}
%%%%%%%%%%%%%%%%%%%%%%%%%%%%%%%%%

\begin{figure}[t]
    \centering
    \includegraphics[width=0.47\textwidth]{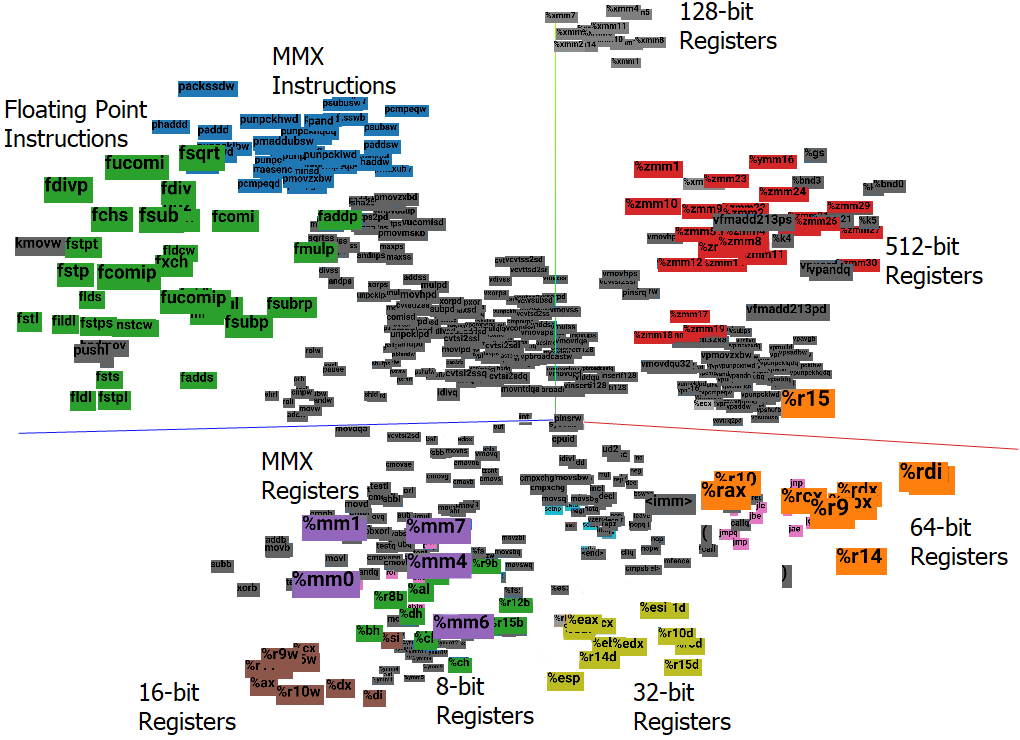}%bert_graph.pdf}
    \caption{3-D visualization for the distribution of instructions and registers after t-SNE is applied to embedding vectors. Similar instructions and registers have the same colors. The unrelated instructions are separated from each other in the three-dimensional space after the pre-training.}
    \label{fig:bert_embedding}
    %\vspace{-0.4cm}
\end{figure}

%
% intro and architecture
%Malicious or vulnerable code detection is a similar task with sentence classification in the NLP literature. 
In an assembly function representation model, the main challenge is to obtain the representation vectors, namely embedding vectors, for each token in a function. Since the skip-gram and RNN-based training models are surpassed by the attention-only models in sentence classification tasks, we introduce FastSpec, which applies a lightweight BERT version.

% training procedure
\subsection{Training Procedures}
We adopt the same training procedures with BERT on assembly functions, namely, \textit{pre-training} and \textit{fine-tuning}.

\subsubsection{Pre-training}
The first procedure is \textit{pre-training}, which includes two unsupervised tasks. The first task follows a similar approach to MaskGAN by masking a portion of tokens in an assembly function. The mask positions are selected from 15\% of the training sequence, and the selected positions are masked and replaced with \texttt{<MASK>} token with 0.80 probability, replaced with a random token with 0.10 probability, or kept as the same token with 0.10 probability. While the masked tokens are predicted based on other tokens' context, the context vectors are obtained by the multi-head self-attention mechanism.

The second task is the next sentence prediction, where the previous sentence is given as input. Since our assembly code data has no paragraph structure where the separate long sequences follow each other, each assembly function is split into pieces with a maximum token size of 50. For the next sentence prediction task, we add \texttt{<CLS>} to each piece. For each piece of function, the following piece is given with the label \texttt{IsNext}, and a random piece of function is given with label \texttt{NotNext}. FastSpec is trained with the self-supervised approach. %Note that the~\texttt{<CLS>} tokens in this phase do not have any meaning about the sequence classes because they are not trained in a supervised manner.

At the end of the \textit{pre-training} procedure, each token is represented by an embedding vector with a size of $H$. Since it is impossible to visualize the high dimensional embedding vectors, we leverage the t-SNE algorithm~\cite{maaten2008t-sne} which maps the embedding vectors to a three-dimensional space as shown in~\autoref{fig:bert_embedding}. We illustrate that the embedding vectors for similar tokens are close to each other in three-dimensional space, as this outcome shows that the embedding vectors are learned efficiently. In~\autoref{fig:bert_embedding}, the registers with different sizes, floating-point instructions, control flow instructions, shift/rotate instructions, set instructions, and MMX instructions/registers are accumulated in separate clusters. The separation among different types of tokens enables achieving a higher success rate in the Spectre gadget detection phase.

\iffalse
\begin{figure}[t]
    \centering
    \includegraphics[width=\columnwidth]{figures/BERT_gadget_added.pdf}
    %\caption{The program with no spectre gadget}
    %\end{subfigure}
    %\centering
    %\begin{subfigure}[t]{.49\columnwidth}
    %\centering
    %\includegraphics[width=\textwidth]{figures/spec_inserted.pdf}
    %\caption{The program with an inserted spectre gadget}
    %\label{fig:pretraining_perplexity}
    %\end{subfigure}
    \caption{(Above) The confidence rates of sliding windows with token size 80 is shown for a benign assembly function. (Below) The inserted Spectre-V1 gadget is detected with 90\% confidence rate in the same function. }
    \label{fig:bert_spec}
    %\vspace{-4mm}
\end{figure}
\fi
\subsubsection{Fine-tuning}
The second procedure is called \textit{fine-tuning}, which corresponds to a supervised sequence classification in FastSpec. This phase enables FastSpec to learn the conceptual differences between Spectre gadgets and general-purpose functions through labeled pieces. The pieces created for the pre-training phase are merged into a single sequence with a maximum of 250 tokens. The disassembled object files, which have more than 250 tokens, split into separate sequences. Each sequence is represented by a single~\texttt{<CLS>} token at the beginning. The benign files are labeled with~\texttt{0}, and the gadget samples are labeled with~\texttt{1} for the supervised classification. Then, the embedding vectors of the corresponding~\texttt{<CLS>} token and position embedding vectors for the first position are summed up. Finally, the resulting vector is fed into the softmax layer, which is fine-tuned with supervised training. The output probabilities of the softmax layer are the predictions on the assembly code sequence.

\subsection{Training Details and Evaluation}\label{sec:training_details}
% data set
%
We combine the assembly data set generated in~\autoref{sec:dataset} and the disassembled Linux libraries to train FastSpec. Although Linux libraries may contain Spectre-V1 gadgets, we assume that the total number of hidden Spectre gadgets is negligible, comparing the data set's total size. Therefore, the model treats those gadgets as noise, which does not affect the performance of FastSpec. In total, a data set of 107 million lines of assembly code is collected, which consists of 370 million tokens after the pre-processing. We separate 80\% of the data set for training and validation, and the remaining 20\% is used for FastSpec evaluation. 
While the same pre-processing phase in~\autoref{sec:gan_details} is implemented, we further merge similar tokens to decrease the total vocabulary size. We replace all labels, immediate values and out-of-vocabulary tokens with \texttt{<label>}, \texttt{<imm>} and \texttt{<UNK>}, respectively. After the pre-processing, the vocabulary size is reduced to 960. 

%For the classification task, we add \texttt{<CLS>} to the beginning of each assembly function.

We choose the number of Transformer blocks as $L=3$, the hidden size as $H=64$, and the number of self-attention heads as $A=2$. We train FastSpec on NVIDIA Titan XP GPU. The \textit{pre-training} phase takes approximately 6 hours, with a sequence length of 50. We further train the positional embeddings for 1 hour with a sequence length of 250. The fine-tuning takes only 20 minutes on the pre-trained model to classify all types of samples in the test data set correctly. Note that the training time is less than previous NLP techniques in the literature since BERT~\cite{devlin2019bert} leverages GPU parallelization significantly. The analysis duration is measured on Intel Xeon CPU E5-2637 v2 @3.50GHz. 

In the evaluation of FastSpec, we obtained 1.3 million true positives and 110 false positives (99.9\% precision rate) in the test data set, demonstrating the high performance of FastSpec. We assume that the false positives are Spectre-like gadgets in Linux libraries, which need to be explored deeply in future work. Moreover, we only have 55 false negatives (99.9\% recall rate), which yield a 0.99 F-1 score on the test data set.

In the next section, we show that FastSpec achieves high performance and extremely fast gadget detection without needing any GPU acceleration since FastSpec is built on a lightweight BERT implementation.

\subsection{Case Study: OpenSSL Analysis}
\label{sec:openssl}
%Although one of the goals of this paper is to create a Spectre-V1 data set that serves as a ground truth for other detection tools and future security research, 
We analyze FastSpec to validate with a separate ground truth data set other than the one we generate in~\autoref{sec:dataset}. The purpose of this analysis is to measure the effect of the covariate shift and robustness of FastSpec against a real-world benchmark. We focus on OpenSSL~v3.0.0 libraries~\cite{openssl}, as it is a popular general-purpose cryptography library in commercial software. We use a subset of functions from RSA, ECDSA, and DSA ciphers in the OpenSSL \textit{speed} benchmark. The function labels are obtained by running the \textit{SpecFuzz} tool, which is a dynamic detection tool to find Spectre-V1 vulnerabilities using fuzzing~\cite{oleksenko2019specfuzz}. The functions in which the \textit{SpecFuzz} tool finds vulnerabilities are labeled as positive, and the remaining ones are labeled as negative. We also exclude the functions without any conditional branch instructions from the positive class and the functions that have a call to them. In total, 4242 functions are extracted from the aforementioned cryptography libraries to analyze with FastSpec. Positive and negative classes include 720 and 2500 functions, respectively.

\begin{figure}[t]
    \centering
    %\vspace{-0.35cm}
    \includegraphics[width=\columnwidth]{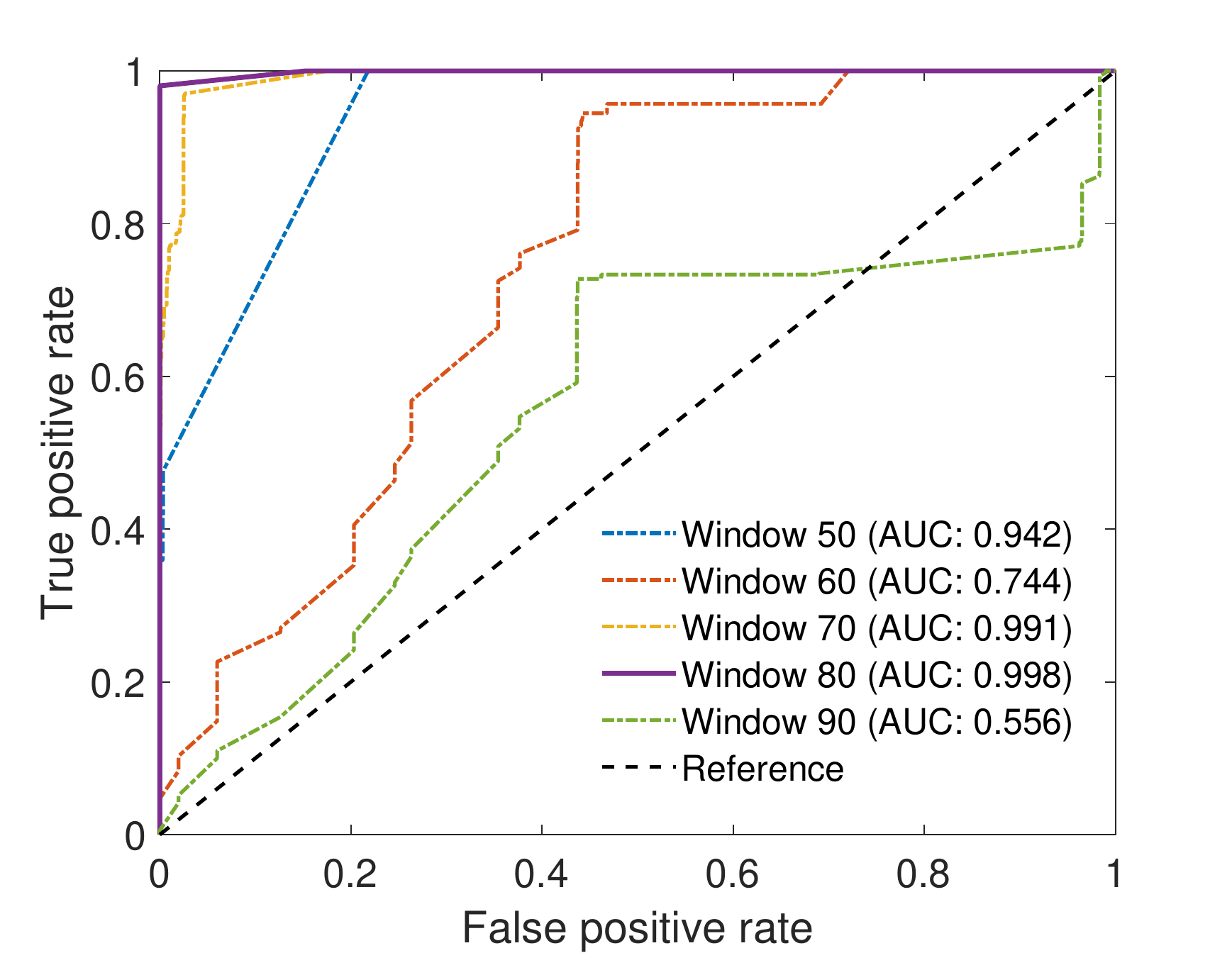}
    \caption{Solid line stands for the ROC curve of FastSpec for Spectre gadget class. Dashed line represents the reference line.}
    \label{fig:roc}
\end{figure}

First, we apply the same pre-processing procedures, as explained in \autoref{sec:training_details} to obtain the tokens. The total number of tokens is more than 4 million. Since the labels are assigned on function-level, we choose the maximum confidence rate that we get among all the sliding windows. The maximum confidence rate is assigned as the prediction of our model for the corresponding input function. In order to find the optimal sliding window size, we scan through the functions with various different window sizes and compare the performances. \autoref{fig:roc} shows that FastSpec achieves the highest performance to detect functions with Spectre-V1 vulnerability when the window size is set to 80 tokens, which corresponds to 0.998 as an area under the curve (AUC) value. The optimal threshold value is found as 0.48, which corresponds to the maximum F-score. The highest F-score is achieved as 0.99, where the false positive rate (benign functions that are mistakenly classified as Spectre gadget) is 0.04\%, and false negative rate (functions that are mistakenly classified as benign) is 2\%. We claim that further analysis of the detected functions by using symbolic execution or taint analysis tools can reduce the number of false negative samples and provide an efficient end-to-end security solution against Spectre-V1 vulnerability.

\subsection{Case Study: Phoronix Test Suite Analysis}\label{sec:phoronix}

The performance comparison between FastSpec and other static analysis tools is evaluated on the Phoronix Test Suite v5.2.1~\cite{phoronix}. For the ground truth, the \textit{SpecFuzz} technique is chosen as the tool that dynamically analyzes the binaries, and exploitable gadgets can be detected with a higher success rate compared to static tools.  The selected benign files have source code since it is required to obtain the assembly files for the \textit{Spectector} tool. The assembly files are generated by compiling the source C code with the \textit{GCC} compiler. On the other hand, the binary files are generated at the test installation; therefore, there is no further processing required before testing the binary files in \textit{oo7}. For FastSpec, the disassembled binary files are given as input. Note that since the larger benchmarks take more time to be analyzed by \textit{oo7}, we preferred small size files to make the comparison with \textit{Spectector} and FastSpec easier.

%\subsubsection{Observations}
\textbf{Timing} The overall timing results for various benchmarks are given in~\autoref{tab:phoronix}. The analysis time of \textit{oo7} and \textit{Spectector} increases drastically with the number of conditional branches since the tools analyze both paths after a conditional branch is encountered. On the other hand, FastSpec analysis time increases linearly with the binary size. We observe that the pre-processing phase takes the major portion in the analysis time of FastSpec while the inference time is in the order of microseconds. We fuzz the Crafty benchmark for 24 hours and other benchmarks for 1 hour using \textit{SpecFuzz} under the default configuration~\footnote{https://github.com/OleksiiOleksenko/SpecFuzz}.

\begin{figure}[t]
    \centering
    \includegraphics[width=1\columnwidth]{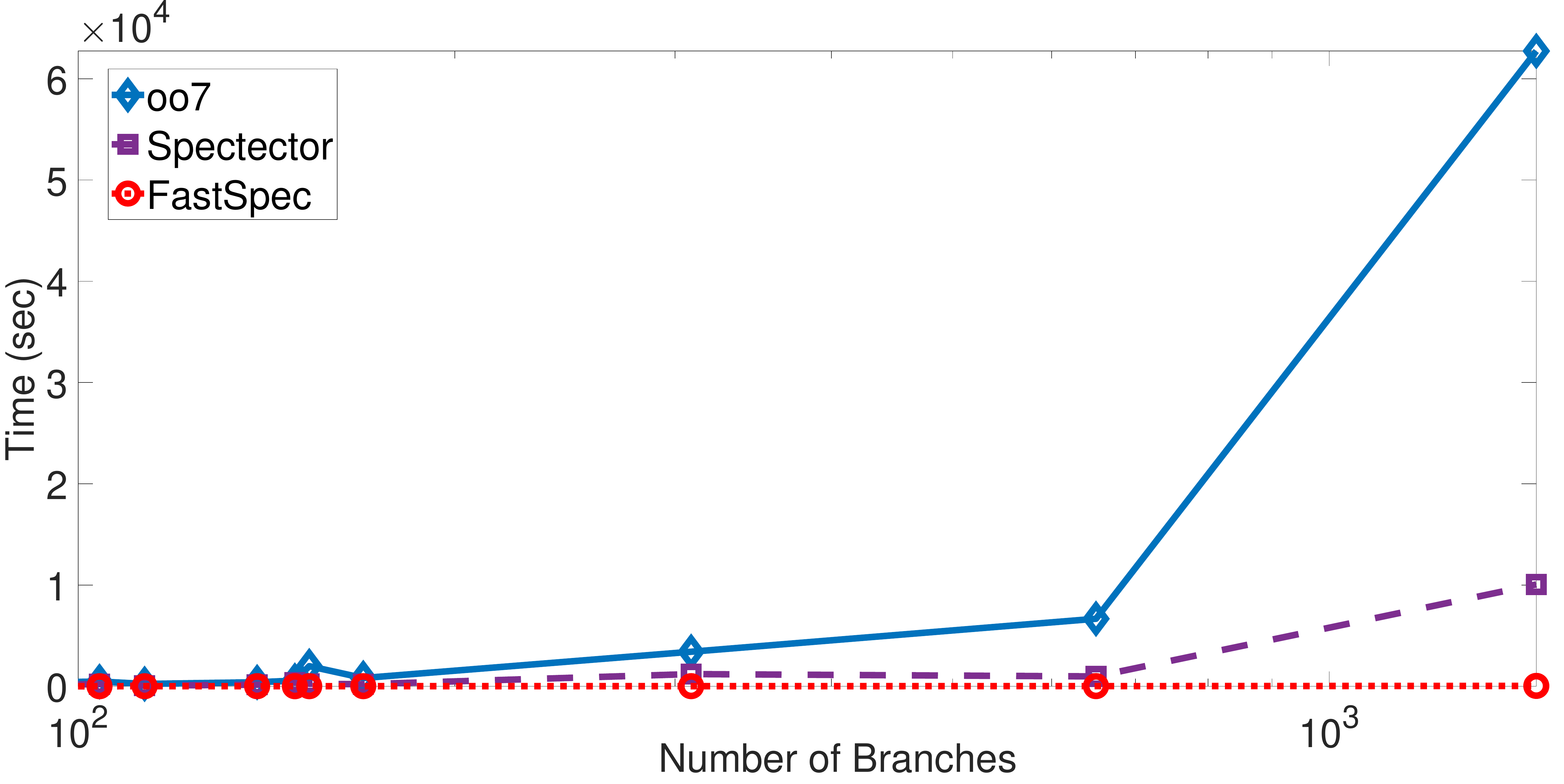}
    \caption{The processing time of FastSpec is independent of the number of branches whereas for Spectector and oo7 the analysis time increases drastically.}
    \label{fig:fastspec_branch}
    %\vspace{-2mm}
\end{figure}

The effect of the increasing number of branches on time consumption is clear in the Crafty and Clomp benchmarks in~\autoref{tab:phoronix}. Even though the Crafty benchmark has only 10,796 branches, \textit{oo7} and \textit{Spectector} analyze the file in more than \textbf{10 days} (the analysis process is terminated after 10 days) and \textbf{2 days}, respectively. In \autoref{fig:fastspec_branch}, we show that both tools are not sufficiently scalable to be used in real-world applications, especially when the files contain thousands of conditional branches. Especially \textit{oo7} shows an exponential behavior because of the forced execution approach, which executes every possible path of the conditional branches.  In contrast, FastSpec analyzes the same Crafty benchmark under 6 minutes, which is a significant improvement.

Note that the Byte benchmark has a higher number of branches than most of the remaining benchmarks. However, it consists of multiple independent files that need to be tested separately, taking less time to analyze in total. Consequently, FastSpec is faster than \textit{oo7} and \textit{Spectector} 455 times and 75 times on average, respectively.

\begin{table*}
\centering
\footnotesize
\caption{Comparison of \textit{oo7}~\cite{wang2018oo7}, Spectector~\cite{guarnieri2020spectector}, and FastSpec on the Phoronix Test Suite. The last column shows that FastSpec is on average 455 times faster than \textit{oo7} and 75 times faster than \textit{Spectector}. (\#CB: Number of conditional branches, \#Fc: Number of functions, \#DFc: Number of detected functions) }
\resizebox{\textwidth}{!}{%
\begin{tabular}{l|c|c|c||c|c|c|c|c|c|c|c|c|c} \toprule
%\begin{tabularx}{\textwidth}{@{}llllllllll@{}}%{C}c@{}}
           & & & & \textbf{SpecFuzz} & \multicolumn{3}{c|}{\textbf{oo7}} & \multicolumn{3}{c|}{\textbf{Spectector}} & \multicolumn{3}{c}{\textbf{FastSpec}} \\
           Benchmark & \thead{Size\\(KB)} & \thead{\#CB} & \#Fc & \thead{\#DFc} & \thead{\\Precision} & \thead{\\Recall}  & \thead{Time \\(sec)} & \thead{\\Precision} & \thead{\\Recall} & \thead{Time \\(sec)} & \thead{\\Precision} & \thead{\\Recall}  & \thead{Time \\(sec)}\\
           \midrule
Byte       & 183.5 &  363   &  83       &7       & 0.70  &\textbf{0.90}   & 400           &1.00   &0.43   & 115       & \textbf{1.00}    &   0.86     &  \textbf{14} \\
%Cachebench & 27.7  &  149   &  18       &NA       & ?     &?      & 556           & ?      &?      & 360       & \textbf{?}    &   \textbf{?}     &  \textbf{5}  \\
Clomp      & 79.4  &  1464  &  45       &1        & 0     &0      & 17.5 hr       & 0.05   &0.9    & 2.8 hr    & \textbf{1.00} &   \textbf{1.00}  &  \textbf{35} \\
Crafty     & 594.8 &  10796 &  207      &44       & \textbf{1.00}  &0.54   & $>$10 day     & 0.60   &\textbf{0.91}   & 48 hr     & 0.23 &   0.80  &  \textbf{315}\\
C-ray      & 27.2  &  139   &  11       &1        & \textbf{1.00} &1.00   & 395           & 0.2    &0.9    & 153       & 0.50 &   \textbf{1.00}  &  \textbf{8}  \\
Ebizzy     & 18.5  &  104   &  6        &3        & 0     &0      & 467           & 0.60   &1.00   & 206       & \textbf{1.00} &   0.33  &  \textbf{3}  \\
Mbw        & 13.2  &  70    &  5        &1        & 0     &0      & 145           & \textbf{0.50}   &1.00   & 34        & 0.33 &   \textbf{1.00}  &  \textbf{2}  \\
M-queens   & 13.4  &  51    &  4        &1        & 1.00  &1.00   & 136           & 0.50   &1.00  & 24        & \textbf{1.00} &   \textbf{1.00}  &  \textbf{2}  \\
Postmark   & 38.0  &  309   &  49       &6        & 1.00  &0.83   & 3409          & 0.43   &0.95   & 1202      & \textbf{1.00} &   \textbf{1.00}  &  \textbf{10} \\
Stream     & 22.0  &  113   &  4        &3        & 0     &0      & 231           & 0      &0      & 63        & \textbf{1.00} &   \textbf{0.66}  &  \textbf{4}  \\
Tiobench   & 36.1  &  169   &  19       &1        & 0     &0      & 813           & 0.25   &0.8    & 201       & \textbf{0.33} &   \textbf{1.00}  &  \textbf{9}  \\
Tscp       & 40.8  &  651   &  38       &13       & 0     &0      & 6667          & 1.00  &0.15   & 972       & \textbf{1.00} &   \textbf{0.92}  &  \textbf{12} \\
%T-test1    & 13.7  &  47    &  3        &NA       & ?     &?      & 99            & ?      &?      & 36        & \textbf{?}    &   \textbf{?}     &  \textbf{3}  \\
Xsbench    & 27.9  &  153   &  32       &1        & \textbf{1.00}  &\textbf{1.00}   & 1985          & 0      &0      & 249       & 0.50 &   0.90  &  \textbf{7}  \\
\midrule
Average    &      \multicolumn{4}{c|}{ }       & 0.47  &0.44  &           & 0.43      &0.67      &        & \textbf{0.74} &   \textbf{0.87}  &   
\\ \bottomrule
\end{tabular}
}
%\vspace{-1cm}
\label{tab:phoronix}
%\end{tabularx}
\end{table*}

\textbf{Baseline Evaluation} The number of gadgets found by the tools varies significantly. While \textit{oo7} and FastSpec report each Spectre gadget in a binary file, \textit{Spectector} outputs whether a function contains a Spectre gadget or not. To be consistent, if a control or data leakage is found in a function, it is reported as a vulnerable function by all three tools. 

The precision and recall rates for \textit{oo7}, \textit{Spectector} and FastSpec are given in~\autoref{tab:phoronix}. The precision is calculated as $TP/(TP+FP)$. TP is the number of overlapping gadgets detected by a tool. FP is the number of functions that are classified as Spectre gadgets mistakenly. The recall value is computed as $TP/(TP+FN)$ where FN is the number of gadgets that are not detected by a tool.

In some cases, \textit{oo7} is not able to track the control flow when the number of function calls increases in a gadget, which yields high false negatives and low recall. Thus, \textit{oo7} suffers from the extraction of complete control flow graph. \textit{Spectector} tends to give more false positives compared to \textit{oo7} and FastSpec. This is because some unsupported instructions are skipped by the tool and the broken Spectre gadgets by specific instructions are still classified as Spectre gadget. On the other hand, FastSpec has low false negatives since all the Spectre gadget patterns are detected with a confidence rate higher than 0.48. When the file size increases, the false positives may increase in parallel. However, these gadgets can be verified with other tools to increase the confidence. As a result, FastSpec scans the functions extremely quicker than other tools without sacrificing the precision and recall rates. Our tool also guarantees the security of the scanned assembly functions by detecting almost all Spectre gadgets with low FN rates. FastSpec outperforms all the compared tools in terms of recall and precision rates by a large margin.

\section{Discussion and Limitations}\label{sec:discussion}
%%%%%%%%%%%%%%%%%%%%%%%%

\subsection{Gadget Verification}%\textbf{Gadget Verification:} 
The gadget verification process in~\autoref{sec:gadget_verification} is implemented in an isolated core since the system interrupts frequently mistrain the targeted branch instructions in the gadgets, which decreases the gadget verification success rate significantly. This situation particularly affects the first step of the verification process where all the inputs are out-of-bounds, and the target branch is not expected to be mistrained. Therefore, there is a need for an isolated environment to run the verification code for Spectre gadgets. Even though the data cache side-channel is used for the secret decoding, other side-channels can be used to decode the secret in a Spectre gadget such as TLB structure. The secret elements in \textit{array1} should be multiplied with a constant to decode the secret into different cache lines or pages. In the base examples~\cite{kocher2018spectre}, the secret elements are multiplied by 512 or 4096. The verification code only selects the Spectre gadgets with these specific multiplicands, which potentially introduces a bias in the data set. Since all multipliers in the Spectre gadgets are represented with the same token, \texttt{<imm>}, our detection tool is not affected by the bias introduced by different multipliers. For instance, in OpenSSL and in Phoronix, we observed that gadgets with different multiplicands are detected by our detection tool. %On the other hand, complex encodings and implicit leakages potentially increase the false negative rate in the detection phase. %Thus, the detection capability of FastSpec is not dependent on the multipliers that are used in the gadgets. 

Our verification codes also focus on more complex leakage snippets in which the secret is not simply leaked with a simple multiplication. We observed that similar control-flow statements and more complex encoding techniques are present among Kocher's examples~\cite{kocher2018spectre} (Examples 10-15). After new gadgets are generated from these examples, we observed that these gadgets can still be detected by our verification code. However, if the leakage mechanism in the gadget is altered significantly, it is likely that the secret in the generated gadget is not recovered during verification. Unfortunately, this introduces a bias in our data set as the diversity of the gadgets is limited. Moreover, our detection tool might not be able to detect more complex gadgets as these gadgets are not included in the training data set. To include more complex gadgets in the data set, the verification code can be changed dynamically by analyzing each generated assembly code, which is left as future work.

\subsection{Scalability and Flexibility} %\textbf{Scalability and Flexibility:}
\textbf{Other Spectre Variants:} 
Since pre-training teaches the general assembly syntax and takes a major part in the training process, our pre-trained FastSpec model can be used after fine-tuning for any assembly code task. The modifications are needed only to Step~1 and Step~4 in ~\autoref{sec:gadget_verification} since we need an initial data set and verification code to build up a larger data set. For Spectre~v1.1~\cite{kiriansky2018speculative}, our verification code can be adapted by adding one more attacker-controlled input to verify whether a speculative load is executed or not. Similarly, the speculatively written value in Spectre~v1.2~\cite{kiriansky2018speculative} can be mapped to cache lines to verify the generated gadgets. For Spectre~v2~\cite{kocher2019spectre}, verification procedure needs to be completely changed as the branch instruction is not a conditional branch anymore. For this purpose, the verification code can be modified to mistrain the indirect jumps with attacker known addresses, and then, the secret bytes in the attacker-controlled function are mapped to separate cache lines. Since Spectre-RSB~\cite{spectrereturns} works in a similar way, except \textit{ret} instruction is targeted, the same verification procedure can be adapted.
Finally, in Spectre~v4~\cite{storetoload}, the verification code can supply attacker-controlled variables to specific registers, and then, speculatively loaded data can be decoded to a shared memory to verify the gadgets.

\textbf{Other Attacks:}
Our approach can detect the target \texttt{SMoTher}-gadgets~\cite{bhattacharyya2019smotherspectre} in the code space. The verification procedure in \autoref{sec:gadget_verification}, specifically Step~4, needs to be changed to analyze port fingerprints. For this purpose, the timing of various instructions that are mapped to certain ports can be measured to detect the leaked secrets as implemented in~\cite{aldaya2019port}. It is highly likely that the verification takes more time for the generated gadgets since we need to collect more timings to distinguish the cases between secret leakage and no secret leakage. In NetSpectre~\cite{schwarz2019netspectre}, there are two types of gadgets. The leak gadget is very similar to Spectre v1 whereas only one bit is transmitted. Hence, the verification procedure can be modified to profile a single cache line instead of 256 cache lines. The transmit gadget is used to leak the secret data over the network and has a different structure than the leak gadget. To detect the transmit gadgets with our verification code, the Thrash+Reload technique can be adapted to measure the timing difference between cached and non-cached accesses over the network. Again, the verification procedure potentially takes more time to analyze the generated gadgets since the secret transmission speed is significantly lower than Spectre V1.
%smotherspectre
%portsmash

\textbf{Other Architectures and Applications:}
Although we limit the scope of this paper to generating and detecting the Spectre-V1 gadgets on x86 assembly code, the use of SpectreGAN and FastSpec can always be extended to other architectures and applications with only mild effort.
% other use cases
Furthermore, specially designed architectures are not needed when pre-trained embedding representations are used~\cite{devlin2019bert}. Therefore, the pre-trained FastSpec model can be used for any other vulnerability detection, cross-architecture code migration, binary clone detection, and many other assembly-level tasks. 

The fuzzing tool increases the diversity of the generated gadgets by introducing variations that are later learned by the FastSpec tool. In addition, the detection tool learns the generic gadget type rather than training on small details. In \autoref{sec:training_details}, the evaluation of FastSpec also shows that the tool can detect the potential Spectre gadgets with a 99.9\% precision rate.

% limitations 
%isolated core
%\paragraph{\textbf{Core Isolation:}}The boundary check tests are implemented in the isolated cores since we observed that if the core is not isolated, other system calls and interrupts mistrain the conditional branches. Therefore, the secret is leaked even though the targeted branch is not trained by the attacker, which causes a wrong interpretation of gadgets in the gadget generation process.

\subsection{Comparison of FastSpec with Other Tools} %\textbf{Comparison with Other Tools:}
The most significant advantage of FastSpec is the capability of detecting Spectre gadgets quicker than other tools. If an instruction is not introduced in the training phase, the instruction is treated as unknown, and it has a slight effect on the accuracy of FastSpec since a large window of instructions is analyzed to decide on the Spectre gadgets. While the unsupported instructions are an important issue for the \textit{Spectector} tool, FastSpec can be deployed to other architectures such as ARM and AMD. While small modifications in the assembly code increase the chance of bypassing other tools, our tool is more robust against small modifications. It is easier to adapt FastSpec to other Spectre variants as the vector representations of assembly instructions can be directly used to train a separate model for the variants. Moreover, over-tainting and under-tainting issues decrease the accuracy of taint-based static analysis techniques. However, FastSpec tracks the registers, instructions, and memory accesses with a vector representation, which makes it more reliable in large-scale projects.

\subsection{Scope and Limitations}
\textbf{Scope:} Our scope is to generate Spectre-V1 gadgets by using mutational fuzzing and SpectreGAN methods as well as to detect potential Spectre gadgets in benign programs by significantly reducing the analysis time.

\textbf{Guarantees:} Our verification methods in Step 4.1 guarantee that the generated Spectre-V1 gadgets leak the secret bytes through cache side-channel attacks. Moreover, the FastSpec tool detects the Spectre gadgets with a high precision and recall rate by identifying the gadget patterns at the assembly level. Possible False Positive outputs do not affect the security guarantee provided by FastSpec. The analysis time is significantly reduced compared to rule-based detection tools.

FastSpec generalizes well, i.e., it can recognize similar patterns that are not in our training dataset. However, it does not provide assurance of coverage (completeness) since FastSpec is not based on hand-written rules or formal analysis. In order to decrease the False Negative rate, the probabilistic threshold is kept low in the case studies. In contrast, while FastSpec does not provide such guarantees, it is much faster and scales to larger code-bases. %Therefore, as all the other static and dynamic vulnerability/malware detection tools, FastSpec also needs to be fine-tuned (further trained) with newly introduced Spectre gadgets to decrease the False Negative rate by learning new hidden patterns. %FastSpec tool is required to be fine-tuned after new Spectre gadgets are introduced. %While FastSpec has a low false negative rate, false positive rate is slightly higher. Since our purpose is to detect as many Spectre gadgets as we can, the threshold value is kept lower, which increases the false positive rate. Thus, the potential Spectre gadgets can be tested again with taint analysis or formal verification methods to increase the chance of detecting exploitable gadgets. 

\textbf{Assembly Code Generation:} %\textbf{Challenges in Assembly Code Generation:}
The challenges faced in the regular text generation with GANs~\cite{yu2017seqgan,fedus2018maskgan} also exist in assembly code generation. One of the challenges is \textit{mode collapse} in the generator models. Although training the model and generating the gadgets with masking help reduce mode collapse, we observed that our generator model still generates some tokens or patterns of tokens repetitively, reducing the quality of the generated samples and compilation and real gadget generation rates.

In regular text generation, even if the position of a token changes in a sequence, the meaning of the sequence may change while it would still be somewhat acceptable. However, if the position of a token in an assembly function changes, it may result in a compilation error because of the incorrect syntax. Even if the generated assembly function has the correct assembly syntax, the function behavior may be completely different from the expected one due to the position of a few instructions and registers.

The fuzzing-based gadget generation technique is based on known gadget examples. Since there are already 15 versions of Spectre-V1, we use these gadgets as the starting point for fuzzing. On the other hand, the available gadgets for other variants are significantly lower compared to Spectre-V1 gadgets. To solve this issue, other detection tools can be used to detect Spectre gadgets in benign programs. Then, new gadgets can be generated with fuzzing technique. We leave the further investigation of generation other Spectre variants as future work.

\textbf{Window Size:} %\textbf{Window Size:} 
Since Transformer architecture has no utilization of recurrent modeling as RNNs do, the maximum sequence length is needed to be set before the training procedures. Therefore, the sliding window size can be set to at most the maximum sequence length. On the other hand, our experiments show that using lower window sizes than maximum sequence length provides more accurate Spectre gadget detection and provides fine-grain information on the sequence. % The further analysis of window size selection is given in~\autoref{sec:window_size}.

%Register clobber
%OOV in SpectreGAN and BERT
%Boundary check-in mutational fuzzing
%BERT'le test ederken GPU gerekiyor mu? Gerekmiyor.
%%%%%%%%%%%%%%%%%%%%%%%%%%%%
\section{Conclusion}\label{sec:conclusion}

This work, for the first time, proposed NLP inspired approaches for Spectre gadget generation and detection. First, we extended our gadget corpus to 1.1 million samples with a mutational fuzzing technique.  We introduced the SpectreGAN tool that achieves a high success rate in creating new Spectre gadgets by automatically learning the structure of gadgets in assembly language. SpectreGAN overcomes the difficulties of training a large assembly language model, an entirely different domain than natural language. We demonstrate 72\% of the compiled code snippets behave as a Spectre gadget, a massive improvement over fuzzing based generation. Furthermore, we show that our generated gadgets span the speculative domain by introducing new instructions and their perturbations, yielding diverse and novel gadgets. The most exciting gadgets are also introduced as new examples of Spectre-V1 gadgets.
Finally, we propose FastSpec, based on BERT-style neural embedding, to detect the hidden Spectre gadgets. We demonstrate that for large binary files, FastSpec is 2 to 3 orders of magnitude faster than \textit{oo7} and \textit{Spectector} while it still detects more gadgets. We also demonstrate the scalability of FastSpec on OpenSSL libraries to detect potential gadgets. %Overall, implementations of NLP/DL techniques in computer security are promising compared to traditional hand-written rule-based approaches.

%%%%%%%%%%%%%%%%%%%%%%%%%%%%%%%
%\input{sections/limitations}


\section*{Acknowledgments}
We are grateful to our anonymous reviewers for their valuable comments, \textit{oo7}'s~\cite{wang2018oo7} author Ivan Gotovchits and \textit{SpecFuzz}'s~\cite{oleksenko2019specfuzz} author Oleksii Oleksenko for helping us running the \textit{oo7} and \textit{SpecFuzz} tools accurately. This work is supported by the National Science Foundation, under grant CNS-1814406, and in part by Intel Corporation.

\bibliographystyle{IEEEtran}
\bibliography{references}

%\appendix
\renewcommand{\thesection}{\Alph{section}.\arabic{section}}
\setcounter{section}{0}

\begin{appendices}
\section{}

\subsection{Assembly Gadget Examples} 

In this section, corresponding assembly gadget of given examples in~\autoref{sec:dataset} are provided.
\label{sec:gadgets_bypass_oo7_spectector}
%\section{Assembly dump of CMOVcc gadget} \label{sec:assembly_cmov}

\begin{lstlisting}[style=ASMstyleCMOVXCHGSET,
                    frame=single,
                    caption={When the C code in~\autoref{lst:cmov} compiled with certain optimizations (gcc 7-4 with O2 enabled), the generated assembly code contains CMOV instruction which fools \textit{oo7}.},
                    label={lis:assembly_cmov},
                    xleftmargin=2em,
                    framexleftmargin=1.5em,
                    basicstyle=\footnotesize]	
victim_function:
.LFB23:
    movl    global_condition(%rip), %eax
    testl   %eax, %eax
    movl    $0, %eax
    cmovne  %rax, %rdi
    movslq  array1_size(%rip), %rax
    cmpq    %rdi, %rax
    jbe     .L1
    leaq    array1(%rip), %rax
    leaq    array2(%rip), %rdx
    movzbl  (%rax,%rdi), %eax
    sall    $12, %eax
    cltq
    movzbl  (%rdx,%rax), %eax
    andb    %al, temp(%rip)
.L1:
    rep ret
\end{lstlisting}

%\section{Assembly dump of XCHG gadget}\label{sec:assembly_xchg}
\begin{lstlisting}[style=ASMstyleCMOVXCHGSET,
                    frame=single,
                    caption={While generating gadgets with mutational fuzzing technique, this code is generated by our algorithm from Kocher's example 3 (using clang-6.0 with 02 optimization). },
                    label={lis:assembly_xchg},
                    xleftmargin=2em,
                    framexleftmargin=1.5em,
                    basicstyle=\footnotesize]	
victim_function:
    xchg    %rdi, %r13
    cmpl    %esp, %esp
    movl    array1_size(%rip), %eax
    shr     $1,	%r11
    cmpq    %rdi, %rax
    jbe     .LBB1_1
    addq    %r13, %r11
    leaq    array1(%rip), %rax
    movzbl  (%rdi,%rax), %edi
    jmp     leakByteNoinlineFunction
.LBB1_1:
    retq
leakByteNoinlineFunction:
    movl    %edi, %eax
    shlq    $9, %rax
    leaq    array2(%rip), %rcx
    movb    (%rax,%rcx), %al
    andb    %al, temp(%rip)
    retq
\end{lstlisting}

%\section{Assembly dump of SETcc gadget}\label{sec:assembly_setcc}
\newpage
\begin{lstlisting}[style=ASMstyleCMOVXCHGSET,
                    frame=single,
                    caption={While generating gadgets with mutational fuzzing technique, this code is generated by our algorithm from Kocher's example 9 (using clang-6.0 with 02 optimization). The \textcolor{red}{seta \%sil} instruction sets the lowest 8-bit of \%rsi register based on a condition which is not detected by \textit{oo7}.},
                    label={lis:assembly_set},
                    xleftmargin=2em,
                    framexleftmargin=1.5em,
                    basicstyle=\footnotesize]	
victim_function:
    seta    %sil
    cmpl    $0, (%rsi)
    je      .LBB0_2
    addl    %r15d, %r12d
    sarq    $1,	%r11
    addb    %sil, %r15b
    movzbl  array1(%rdi), %eax
    ja      .L1324337
    testw   %r10w, %ax
    shlq    $12, %rax
    nop
    movb    array2(%rax), %al
.L1324337:
    andb    %al, temp(%rip)
.LBB0_2:
    retq
\end{lstlisting}
%\vspace{4cm}

%\section{Novel Gadgets from Spectector Analysis}\label{sec:spectector_example}

\subsection{Instructions and registers inserted randomly in the fuzzing technique} \label{sec:instructions_inserted_by_fuzzing}
\newcolumntype{s}{>{\hsize=0.5\hsize}X}
\newcolumntype{v}{>{\hsize=1.2\hsize}X}
\begin{table}[h]%[t!]
\caption{Instructions and registers inserted randomly in the fuzzing technique.
%In the \textit{Iteration 2}, \texttt{CMOVcc} instruction is removed from \textit{Iteration 1}. In the \textit{Iteration 3}, \texttt{XCHG} instruction is removed from the \textit{Iteration 2}}
}
%\small
\begin{tabularx}{\columnwidth}{X|X|X|v|X|s}
\hline \toprule
\multicolumn{6}{c}{\textbf{Instructions}}  \\ \hline
add    & cmovll & jns     & movzbl      & ror   & subl   \\
addb   & cmp    & js      & movzwl      & sall  & subq   \\
addl   & cmpb   & lea     & mul         & salq  & test   \\
addpd  & cmpl   & leal    & nop         & sarq  & testb  \\
addq   & cmpq   & leaq    & not         & sar   & testl  \\
andb   & imul   & lock    & notq        & sal   & testq  \\
andl   & incq   & mov     & or          & sbbl  & testw  \\
andq   & ja     & movapd  & orl         & sbbq  & xchg   \\
call   & jae    & movaps  & orq         & seta  & xor    \\
callq  & jbe    & movb    & pop         & setae & xorb   \\
cmova  & je     & movd    & popq        & sete  & xorl   \\
cmovaeq& jg     & movdqa  & prefetcht0  & shll  & xorq   \\
cmovbe & jle    & movl    & prefetcht1  & shlq  & lfence \\
cmovbq & jmp    & movq    & push        & shr   & sfence \\
cmovl  & jmpq   & movslq  & pushq       & sub   & mfence \\
cmovle & jne    & movss   & rol         & subb  &        \\
\toprule
\multicolumn{6}{c}{\textbf{Registers}} \\ \hline
rax & eax  & ax   & al   & xmm0  & ymm0  \\
rbx & ebx  & bx   & bl   & xmm1  & ymm1  \\
rcx & ecx  & cx   & cl   & xmm2  & ymm2  \\
rdx & edx  & dx   & dl   & xmm3  & ymm3  \\
rsp & esp  & sp   & spl  & xmm4  & ymm4  \\
rbp & ebp  & bp   & bpl  & xmm5  & ymm5  \\
rsi & esi  & si   & sil  & xmm6  & ymm6  \\
rdi & edi  & di   & dil  & xmm7  & ymm7  \\
r8  & r8d  & r8w  & r8b  & xmm8  & ymm8  \\
r9  & r9d  & r9w  & r9b  & xmm9  & ymm9  \\
r10 & r10d & r10w & r10b & xmm10 & ymm10 \\
r11 & r11d & r11w & r11b & xmm11 & ymm11 \\
r12 & r12d & r12w & r12b & xmm12 & ymm12 \\
r13 & r13d & r13w & r13b & xmm13 & ymm13 \\
r14 & r14d & r14w & r14b & xmm14 & ymm14 \\
r15 & r15d & r15w & r15b & xmm15 & ymm15 \\
\bottomrule
\end{tabularx}

\label{tab:inst_reg}
%\vspace{-7mm}
\end{table}

\end{appendices}

\end{document}